\pdfoutput=1

\documentclass[11pt]{article}
\usepackage{booktabs}
\usepackage[preprint]{acl}
\usepackage{enumitem}
\usepackage{tabularx}
\usepackage{tcolorbox}
\usepackage{multirow}
\usepackage{adjustbox}
\tcbuselibrary{listingsutf8}
\usepackage{xcolor}
\usepackage{times}
\usepackage{latexsym}
\usepackage{amsmath}
\usepackage{pifont}
\usepackage{cleveref}
\newcommand{\tick}{\ding{51}}
\newcommand{\cross}{\ding{55}} 
\newcolumntype{Y}{>{\raggedright\arraybackslash}X}
\newcolumntype{C}{>{\centering\arraybackslash}m{2em}}

\usepackage[T1]{fontenc}

\usepackage[utf8]{inputenc}

\usepackage{microtype}

\usepackage{inconsolata}

\usepackage{graphicx}
\newcommand*\samethanks[1][\value{footnote}]{\footnotemark[#1]}
\title{Measuring Fine-Grained Negotiation Tactics of Humans and LLMs in Diplomacy}

\author{
Wenkai Li \thanks{Equal contribution.} ~~~
Lynnette Hui Xian Ng \samethanks ~~~
Andy Liu ~~~
Daniel Fried \vspace{0.2em} \\
~Carnegie Mellon University\\
~\texttt{\{wenkail, lynnetteng\}@andrew.cmu.edu}
}

\begin{document}
\maketitle

\begin{abstract}
The study of negotiation styles dates back to Aristotle's ethos-pathos-logos rhetoric. Prior efforts primarily studied the success of negotiation agents. Here, we shift the focus towards the styles of negotiation strategies.
Our focus is the strategic dialogue board game Diplomacy, which affords rich natural language negotiation and measures of game success.
We used LLM-as-a-judge to annotate a large human-human set of Diplomacy games for fine-grained negotiation tactics from a sociologically-grounded taxonomy. Using a combination of the It Takes Two and WebDiplomacy datasets, we demonstrate the reliability of our LLM-as-a-Judge framework and show strong correlations between negotiation features and success in the Diplomacy setting. Lastly, we investigate the differences between LLM and human negotiation strategies and show that fine-tuning can steer LLM agents toward more human-like negotiation behaviors. 
\end{abstract}




\begin{figure*}[htbp]
    \centering
    \includegraphics[width=0.95\textwidth]{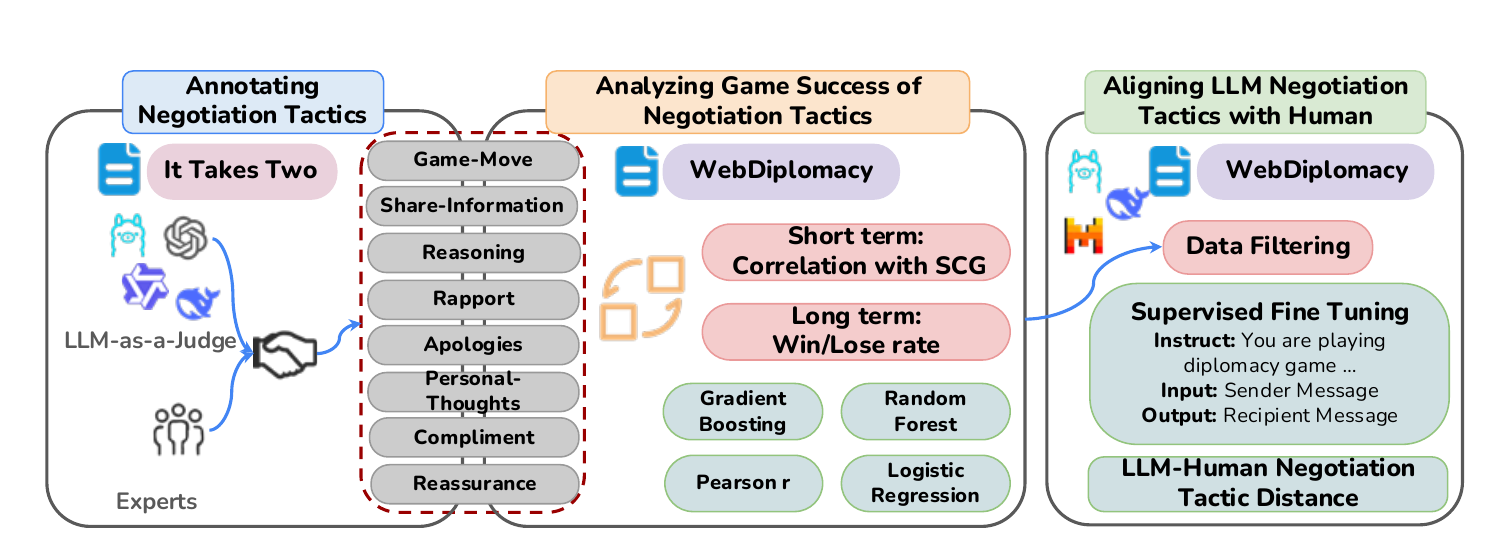}
    \caption{Methodology Overview: Our pipeline consists of three stages: \textbf{(1) Reliable tactic annotation.} We first annotate negotiation tactics with an LLM-as-a-Judge and validate its reliability on the It Takes Two dataset by computing agreement with expert annotators. \textbf{(2) Linking tactics to outcomes.} Using real human communications and game logs from WebDiplomacy, we study how annotated negotiation tactics relate to performance, analyzing short-term correlations and long-term win/loss outcomes. \textbf{(3) Aligning LLMs with humans.} We do supervised finetuning on filtered WebDiplomacy interactions to align LLM negotiation style with human tactics and quantify the LLM–human tactic distance.}
    \label{fig:overview_figure}
    \vspace{-1em}
\end{figure*}

\section{Introduction}
Negotiation has long been studied as both a science and an art, dating back to Aristotle's three modes of rhetoric: Ethos appeals to credibility; Pathos appeals to emotions; and Logos appeals to logic~\cite{kennedy1993aristotle}. How an argument is presented can be as crucial as what is being said; the strategy a negotiator adopts can profoundly affect the outcome of a negotiation.

A growing body of work in NLP and AI has focused on developing agents with strong negotiation abilities. NLP systems have demonstrated impressive negotiation capabilities, including in the strategic negotiation board game Diplomacy~\cite{meta2022human} as well as engaging in multi-issue bargaining%
~\citep{lau2008mining,lewicki2011essentials,lewis2017dealdealendtoendlearning,he2018decoupling}. 

However, most evaluations of AI negotiation agents emphasize objective metrics like win rates, efficiency of the deal, or the balance of concessions \citep{meta2022human, kwon2024llmseffectivenegotiatorssystematic, bianchi2024llmsnegotiatenegotiationarenaplatform, fu2023improvinglanguagemodelnegotiation}. Less focus has been placed in the understanding of the tactics (i.e., rhetoric, tone) models employed in negotiation. The tactics negotiation agents use -- cooperative or combative, persuasive or dismissive -- affect receiver perception, and the agent's effectiveness and reception \citep{chawla2021towards, chawla2022opponent, mell2019likeability, kwon2024llmseffectivenegotiatorssystematic}. Prior efforts to study negotiation style adopted ad-hoc definitions that are insufficiently grounded in past negotiation theory, making it difficult to compare negotiation tactics across studies or to link observed negotiation behaviors.

In this paper, we profile the distribution and impact of fine-grained negotiation tactics through a sociologically grounded framework, using Diplomacy as a testbed. We used two datasets of bilateral human-human dialogues: (1) the \textbf{It Takes Two} dataset, which contains Diplomacy games collected by \citet{peskov2020takes} and annotated for negotiation tactics by \citet{jaidka2023takesnegotiatemodelingsocial}, and (2) a \textbf{WebDiplomacy} dataset taken from the large scale corpus of online Diplomacy games used by ~\citet{meta2022human}. Details of both datasets are presented in \Cref{appn:datasets_details}. Unlike some past work that solely analyzes LLM-LLM negotiations within synthetic scenarios \citep{tang2025dsgbenchdiversestrategicgame, kwon2024llmseffectivenegotiatorssystematic, bianchi2024llmsnegotiatenegotiationarenaplatform}, we use these naturally occuring datasets to ground our negotiation style analysis and development in human gameplay, before applying it to analyze LLMs.
We study the following Research Questions:
\begin{itemize}
    \item \textbf{RQ1:} How can we annotate negotiation tactics to study human negotiation behavior at scale? We develop an LLM-as-a-judge pipeline for efficient and reliable annotations.
    \item \textbf{RQ2: } Do negotiation styles affect game success? We apply regression and predictive modeling to study how styles affect game success in the large-scale WebDiplomacy dataset.
    \item \textbf{RQ3:} What are the differences in negotiation styles between LLMs and Humans? We prompt LLMs with game contexts from the WebDiplomacy dataset and evaluated the negotiation style distribution in comparison to human messages. 
    \item \textbf{RQ4:} Can we steer LLMs to use similar negotiation tactics as humans? We fine-tune LLMs with human data from WebDiplomacy dataset to match the negotiation tactics.
\end{itemize}

\begin{table*}[ht]
\centering
{\small
\begin{tabularx}{\linewidth}{l l Y}
\toprule
Rhetoric & Negotiation Tactic & Definition \\
\midrule
Ethos & Game-Move & Plans, thoughts and goals about a Diplomacy move \\
Ethos & Share-Information & Messages about the history of or information gained about another player's move (except the speaker's and recipient's) \\
Logos & Reasoning & Speculative reasoning, justification of past or future moves  \\
Pathos & Rapport & Build trust and mutual understanding between speaker and recipient \\
Pathos & Apologies & Expressions of regrets or remorse about past moves  \\
Pathos & Personal-Thoughts & Messages that reflect the speaker's opinions or feelings \\
Pathos & Compliment & Positive messages about the recipient or recipient's moves \\
Pathos & Reassurance & Supportive messages about the recipient's game position  \\
\bottomrule
\end{tabularx}
}
\caption{Taxonomy of Negotiation Tactics and Definitions (adapted from \citet{jaidka2023takesnegotiatemodelingsocial})}
\label{tab:negotiation_tactics}
\end{table*}

\section{Related Work}

\paragraph{Diplomacy}
Diplomacy is a strategic board game that requires complex negotiation to form alliances. Seven players aim to control a majority of 34 supply centers on a map of Europe by coordinating the movement of their military units. While Diplomacy is a zero-sum game, players must negotiate strategic coalitions to support their own plans or counteract the moves of other players. 
Bilateral negotiations are held in private and do not bind future moves, meaning that building long-term trust can be critical to game success. 
See \Cref{appn:diplomacy_description} for a more detailed description of the game. 

Diplomacy game dialogue has been used to study perceptions of trust, deception and persuasion, and perceptions of lies \citep{niculae2015linguistic,peskov2020takes,ahuja2022using,ng2025improving,wongkamjan2024more,wongkamjan2025should}.
The game has also been an essential testbed for assessing LLM-powered strategic reasoning \citep{paquette2019pressdiplomacymodelingmultiagent, gray2021humanlevelperformancenopressdiplomacy,bakhtin2022masteringgamenopressdiplomacy}.
Meta's CICERO system successfully combined an LLM-based dialogue system with strategic reasoning, competing successfully against human opponents \citep{meta2022human}. 
However, many of these LLM-based works emphasize endgame outcomes, leaving negotiation dialogue dynamics relatively underexplored. 

\paragraph{LLM as negotiators} Research evaluating LLM negotiation capabilities spans diverse domains: games, finance, law, and business~\citep{kwon2024llmseffectivenegotiatorssystematic, bianchi2024llmsnegotiatenegotiationarenaplatform, fu2023improvinglanguagemodelnegotiation, noh2024llmspersonalitiesmultiissuenegotiation}. \citet{kwon2024llmseffectivenegotiatorssystematic} systematically assessed LLM performance on 35 negotiation tasks, noting GPT‑4's strength but its struggle with subjective judgment and strategic adaptability. \citet{bianchi2024llmsnegotiatenegotiationarenaplatform} presented NegotiationArena, revealing how LLMs develop strategic and irrational tactics in negotiation exchanges.

Stylistic linguistic features in dialog can reflect power and influence \cite{niculae2015linguistic}, which suggests that agents that linguistically adapt can gain a social or persuasive edge. LLM and human negotiation behavior can be rather different \cite{wongkamjan2024more}, which leads onto investigations to shift LLM behavior for better alignment and authority.
Prompt-based interventions and fine-tuning can be effective to shift LLM behavior. \citet{noh2024llmspersonalitiesmultiissuenegotiation} found that personality-driven prompts can shift LLM negotiation behavior from cooperative to adversarial without retraining. Reinforcement-learning-inspired methods, such as self-play with feedback, have also been shown to improve negotiation success~\citep{lewis2017dealdealendtoendlearning,fu2023improvinglanguagemodelnegotiation, chen2023effective,liao2024efficacy}. Such works highlight both the potential and limitations of LLM negotiation capabilities, motivating our studies of how LLM agents can use fine-grained negotiation tactics and how far they are from humans in a strategic Diplomacy environment. Building on these past works, we perform a large-scale analysis of negotiation strategies, enabled by LLM-as-a-judge, and examine the relationship between these strategies and success in the game.

\begin{figure*}[htbp]
    \centering
    \includegraphics[width=1\textwidth]{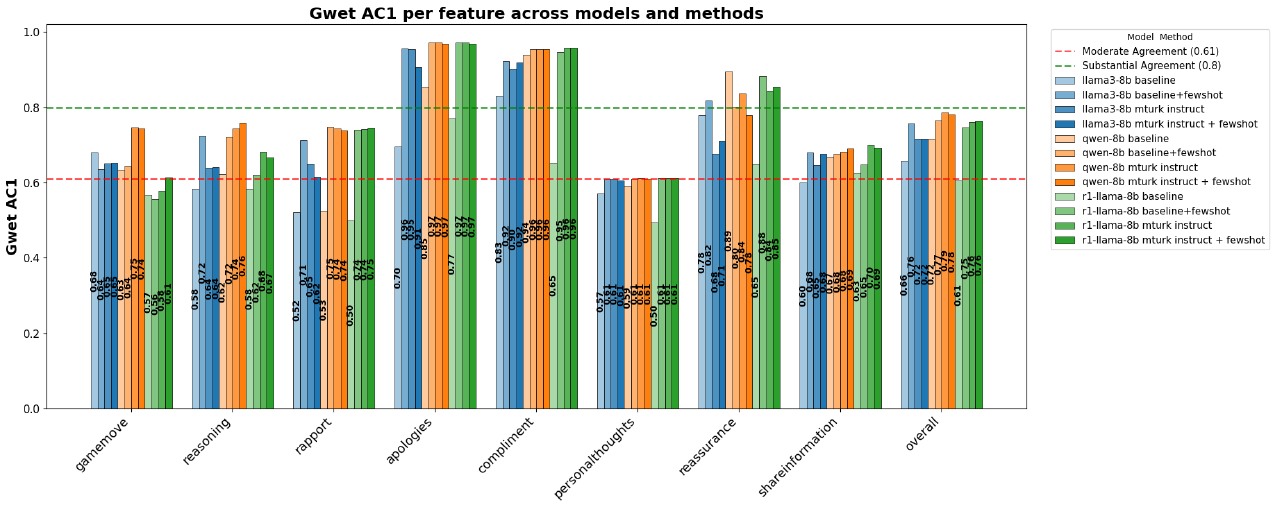}
    \caption{Gwet’s~AC1 agreement scores per negotiation tactic across models and prompting methods when compared against the expert gold standard. The dashed red line indicates the threshold for moderate agreement ($\text{AC1}=0.61$), while the dashed green line indicates substantial agreement ($\text{AC1}=0.8$).}
    \label{fig:gwet_ac1_by_feature}
\end{figure*}

\section{Negotiation Tactics Analysis}
\label{sec:annotations}

We analyze Diplomacy dialogue using a taxonomy of fine-grained negotiation tactics adapted from \citet{jaidka2023takesnegotiatemodelingsocial}. (See \Cref{appn:diplomacy_description} for a detailed description of the Diplomacy game.)  This taxonomy is based on the Ethos-Pathos-Logos rhetoric, and breaks down negotiation into eight tactics, each serving a psychological and strategic function that contributes to negotiation effectiveness. The tactics, definitions are listed in \autoref{tab:negotiation_tactics}, and their sociological groundings in \autoref{tab:negotiation_tactics_sociological}.
We then correlate the presence of each negotiation tactic with game success in human-human games. We also evaluate LLMs on their use of these strategies in self-play, and compare them to humans. \autoref{fig:overview_figure} illustrates our methodology.

Past work annotated messages with the negotiation strategies using Amazon Mechanical Turk workers ~\citep{jaidka2023takesnegotiatemodelingsocial}. However, the nuanced nature of the task resulted in differing interpretations among the annotators, and therefore inconsistent crowd-sourced labels \citep{ng2025improving}. 

Therefore, we develop an LLM-as-a-judge pipeline as a scalable and reliable approach to annotate the messages. We prompted models to perform binary classifications (presence or absence of each tactic) in a single prompt on a subset of messages ($n=128$) from the \textbf{It takes two} dataset~\cite{peskov2020takes,jaidka2023takesnegotiatemodelingsocial}. The models were: \texttt{LLaMA3.1-8B-Instruct} \citep{dubey2024llama}, \texttt{Qwen-3-8B} \citep{qwen3technicalreport}, and \texttt{r1-distilled-LLaMA-8B} \citep{deepseekai2025deepseekr1incentivizingreasoningcapability}. 
Details prompts are provided in \Cref{appendix:prompt_templates}:
\begin{itemize}
    \item \textbf{Baseline (Zero-shot)}: A direct prompt asking the model to judge each of the eight tactics without any instruction or examples.
    \item \textbf{Few-Shot}: Providing some positive example of each tactic from expert annotators. In total, eight examples were provided.
    \item \textbf{Instructions}: The original task description used for crowd workers from \citet{jaidka2023takesnegotiatemodelingsocial}, which included definitions and decision rules for each feature.
    \item \textbf{Instructions + Few-shot}: A hybrid prompt that included both the instruction template and the few-shot examples.
\end{itemize}

We compare the LLM annotations with expert annotations. Three expert annotators (authors of this paper) annotated a subset of randomly selected $n=128$ dialogue messages, guided by the same instructions as the crowd-sourced annotators (see \Cref{fig:mturk_instruct_1,fig:mturk_instruct_2,fig:mturk_instruct_3}). 
Because our annotation label distribution is notably imbalanced across categories, we report Gwet’s AC1, a chance-corrected agreement statistic that is more robust than Fleiss-kappa agreement to prevalence and marginal asymmetry (More explanation is in \Cref{appn:llm_as_a_judge_details}). Agreement among experts achieved substantial reliability (overall mean AC1 = $0.678$), supporting the feasibility of a gold-standard label set. Comparing LLM annotations against the expert labels yields moderate agreement on average, whereas crowd-sourced annotators exhibit markedly lower agreement with experts (below the moderate threshold), indicating that LLM-as-a-judge is a more dependable path for scaling annotation than crowd labels alone.

\Cref{fig:gwet_ac1_by_feature} reports AC1 scores between each LLM and the expert annotations, broken down by conversational feature:
\begin{itemize}[leftmargin=1em, topsep=1pt, itemsep=1pt, parsep=0pt]
    \item \textbf{Prompting scheme dominates size:}  
        Across models, \textbf{baseline} prompts sit well below the substantial-agreement band. Adding \textbf{instruction} prompting consistently lifts AC1, and adding \textbf{few-shot} examples produces the largest gains—often pushing scores above the \textbf{moderate} (AC1 \(=0.61\)) line and, for several features, into the \textbf{substantial} (AC1 \(=0.80\)) range.
    \item \textbf{Qwen-8B is the most reliable judge:}  
        With \textbf{instructions + few-shot}, \textsc{Qwen-8B} attains the highest mean AC1 across features, edging out \textsc{Llama-3-8B} and the distilled \textsc{r1-Llama-3-8B}. \textsc{Qwen-8B} also leads on a majority of individual features.
    \item \textbf{Feature-level difficulty varies:}
        Socially straightforward tactics such as \textsc{apologies} and \textsc{compliments} achieve high agreement. \textsc{reassurance} typically lands around the substantial threshold. In contrast, \textsc{personal-thoughts} remains challenging (near the moderate band), while \textsc{rapport}, \textsc{reasoning}, \textsc{game-move}, and \textsc{share-information} fall in the midrange.
\end{itemize}

\begin{table*}[htbp]
\centering
\resizebox{\linewidth}{!}{%
\begin{tabular}{lcccccccc}
\toprule
tactic & point\_biserial\_r & p\_pb & spearman\_r & p\_sp & cohen\_d & rank\_biserial\_r \\
\midrule
1. Game-Move & 0.236 & $<$1e$-$6 & 0.362 & $<$1e$-$6 & 0.278  & -0.137 \\
2. Reasoning & 0.180 & $<$1e$-$6 & 0.296 & $<$1e$-$6 & 0.341 & -0.196 \\
3. Rapport & 0.200 & $<$1e$-$6 & 0.290 & $<$1e$-$6 & 0.348 &  -0.203 \\
4. Apologies & 0.179 & $<$1e$-$6 & 0.234 & $<$1e$-$6 & 0.382 & -0.227 \\
5. Compliment & 0.152 & $<$1e$-$6 & 0.216 & $<$1e$-$6 & 0.376 & -0.221 \\
6. Personal-Thoughts & 0.127 & $<$1e$-$6 & 0.171 & $<$1e$-$6 & 0.364 & -0.215 \\
7. Reassurance & 0.147 & $<$1e$-$6 & 0.234 & $<$1e$-$6 & 0.309 & -0.188 \\
8. Share-Information & 0.182 & $<$1e$-$6 & 0.293 & $<$1e$-$6 & 0.301 & -0.172 \\
\bottomrule
\end{tabular}
}
\caption{Correlation and Effect Size  between negotiation tactics and yearly SCG. All tactics show statistically significant positive correlations with SCG ($p<1e{-6}$), supporting the hypothesis that both tactical reasoning and socio-emotional strategies contribute meaningfully to short-term success.}
\label{tab:pearson_correlation_between_features_scg}
\end{table*}

\section{Analyzing Negotiation Style's Effect on Game Success}
In this section, we investigate whether negotiation tactics affect game success in the WebDiplomacy dataset (see more detailed description in \Cref{appn:datasets_details}). We use \textsc{Qwen3‑8B} to annotate all messages or the presence of each of the eight negotiation strategies. The labels were then aggregated at the phase level per player, yielding both a binary indicator and a count (frequency of occurrences) for each feature in each player-phase.

Game success was measured using two metrics: short-term success with Supply Center Gain (SCG) gaining and long-term success with final winning.

\subsection{Short Term Success}
\label{sec:shorttermsuccess}
We define the player's Supply Center Gain (SCG) as a measurement of success, using meta-data from WebDiplomacy. SCG is the net change of supply centers controlled at the end of each game year.
The SCG per player per year is a continuous outcome variable that was positive if the player gained centers, negative if centers were lost, and zero if the number of centers remained unchanged.

\begin{figure}[htbp]
    \centering
    \includegraphics[width=\columnwidth]{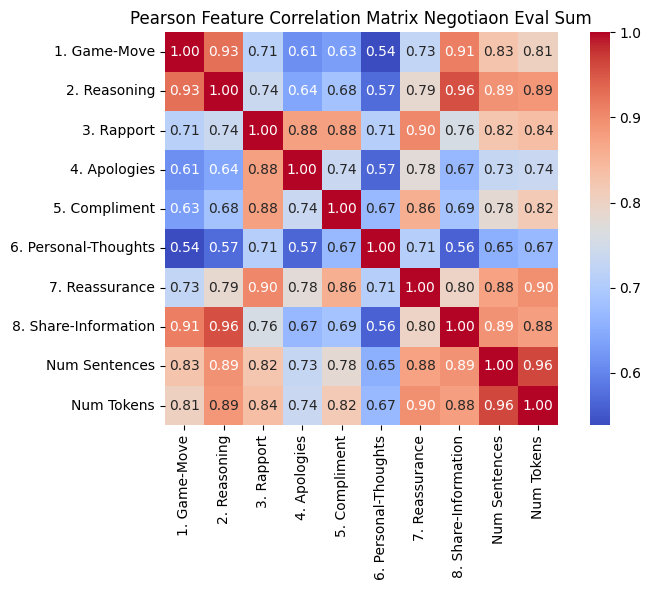}
    \caption{Correlation between annotated negotiation features and supply center gain.}
    \label{fig:feature_correlation}
\end{figure}

\paragraph{Correlation Analysis}
We first examine simple correlations between negotiation strategies and SCGs at the phase level. Since the measurement of supply centers occurred every game year, we considered the collective sum of the presence of features for each year for each power. \Cref{fig:feature_correlation} shows the correlation between each feature. We controlled for length, as we found that the number of sentences sent per player-phase was strongly correlated with each negotiation tactic (Num Sentences and Num Tokens correlated by $\ge\!0.83$).

We computed the Pearson’s $r$ as a point-biserial correlation \citep{benesty2009pearson} between negotiation tactics and SCGs. $r$ quantifies the strength and direction of linear relationships between continuous features and outcomes, making it well-suited to analyze how the frequency of each tactic relates to SCGs. This associates whether players who used a given tactic during a phase tended to gain more supply centers at the end of the phase. As presented in \autoref{tab:pearson_correlation_between_features_scg}, all eight stylistic dimensions show statistically significant positive Pearson correlations with yearly supply-center gain ($p<10^{-6}$ after a Bonferroni correction).

The strongest linear association arose from the tactical \textsc{Game-Move} ($r=.24$), demonstrating that tactical discussion of moves and strategies enhances negotiation outcomes by reducing uncertainty \cite{bazerman1993negotiating}. The next strongest linear associations were interpersonal \textsc{Rapport} ($r=.20$), mirroring how rapport-building significantly improves negotiations from increased trust \cite{drolet2000rapport}, and analytical \textsc{Reasoning} ($r=.18$), supporting how logical arguments are most effective in strategy games \cite{petty1986elaboration}. Social-politeness markers such as \textsc{Apologies}, \textsc{Compliment}, and \textsc{Reassurance} still had positive associations, albeit with smaller effects ($.13\le r\le .18$), which reflects how social behaviors reduces resistance and facilitate cooperation \cite{brown1987politeness}. Information exchange (\textsc{Share-Information}, $r=.18$) sits mid-table, suggesting that while this strategy can improve outcomes, it also creates vulnerability in revealing the player's position \cite{galinsky2001first}. 

For robustness analysis, we extended the inquiry to a frequency-adjusted regression (see \autoref{sec:appendix_ols}).
These correlational analyses (binary presence and frequency-adjusted) demonstrate that the taxonomy of negotiation tactics are correlated with short-term outcomes, highlighting the robustness of the taxonomy and the importance of fine-grained negotiation tactic analysis.

\subsection{Predictive Modeling}
To move beyond univariate correlations and gain a more comprehensive understanding of how negotiation strategies relate to success in Diplomacy, we used predictive modeling analysis with machine learning (ML) methods. Predictive modeling analysis provides stronger information about the explanatory power of negotiation tactics. If negotiation tactics matter for performance, then a predictive model trained on the negotiation features should be able to accurately forecast player success.
For robustness, we also used an Ordinary Least Squares regression to validate the predictive nature of the negotiation tactics (more detailed explanations are in \Cref{sec:appendix_predictive}). The OLS model allows us to uncover the relative importance among tactics in their contribution to the player's game success.

Using ML prediction techniques, we evaluate the power of negotiation features for short-term success. We treated SCG as the prediction target for a suite of supervised machine learning models: Logistic Regression, Random Forest, and Gradient Boosting. Model inputs were either per-phase negotiation feature counts or their standardized aggregated frequencies across the game. Model training and hyperparameter optimization were performed via cross-validation, with evaluation on a held-out test set using metrics of accuracy, F1-score, and ROC-AUC. All three tested models hovered around $\mathbf{61\%}$ accuracy and $\mathbf{65\%}$ ROC‑AUC, substantially above the majority baseline (50\%) (see \autoref{tab:logic_regression_model_prediction}). We analyzed feature importance scores to interpret model decisions. Our integrated regression and prediction framework allows us to identify not only which negotiation behaviors correlate with but also are predictive of player success. 

The Gradient Boosted classifier was the most accurate. Its top-20 feature importances (see \Cref{fig:feature_importance}) closely echoed the OLS findings: 
\begin{itemize}[leftmargin=1em, topsep=1pt, itemsep=1pt, parsep=0pt]
    \item \textbf{Game‑Move} dominated predictability (18.7\%), reinforcing its role as the single best indicator of positive SCG. The dominance of Game-Move aligns with costly signaling theory because these communications are the most costly form of signaling and difficult to fake -- sharing specific tactical information requires deep analysis and carries strategic risks, making the signals reliable indicators of genuine cooperation \cite{przepiorka2017signaling}.
    \item \textbf{Rapport} (11.6\%) and \textbf{Reassurance} (5.6\%) followed, showing that well‑timed socio‑emotional cues acts as social exchanges \cite{blau2017exchange}, which therefore add predictive value. 
    \item Length effects appear both directly (num\_tokens, 3.7\%) and via interactions (e.g.\ \textbf{Game-Move\,$\times$\,Share‑Information}), underlining how longer, more detailed messages serve as heuristic indicators of sender effort and seriousness and results in deeper evaluation of proposals \cite{petty1986elaboration}.
\end{itemize}

\subsection{Long Term Success}
We represent long-term success by the eventual game outcome (win or loss). We compared the breakdown of negotiation tactics between the eventual winners and losers, by comparing the average frequency each negotiation strategy was used by winning players in comparison to the losers. For each game, we calculate the average rate of each negotiation style per phase for the winner and a randomly sampled loser. To account for differences in total message volume, we perform normalization per year. We then compare these average feature frequencies between the winners and losers. Next, to isolate the effect of negotiation tactics, regardless of the strength of the player’s position (reflected by their supply center count), we condition the frequency on the supply center counts at each phase. This controlled for the cumulative advantages and opportunities that players with more centers have, and allowed better discernment on whether winners exhibited distinct negotiation.
This long-term analysis focuses on the differences where communication behavior correlates with ultimate success, offering insight into the characteristics of winning sets of negotiation tactics.

\begin{figure}[htbp]
    \centering
    \includegraphics[width=1\columnwidth]{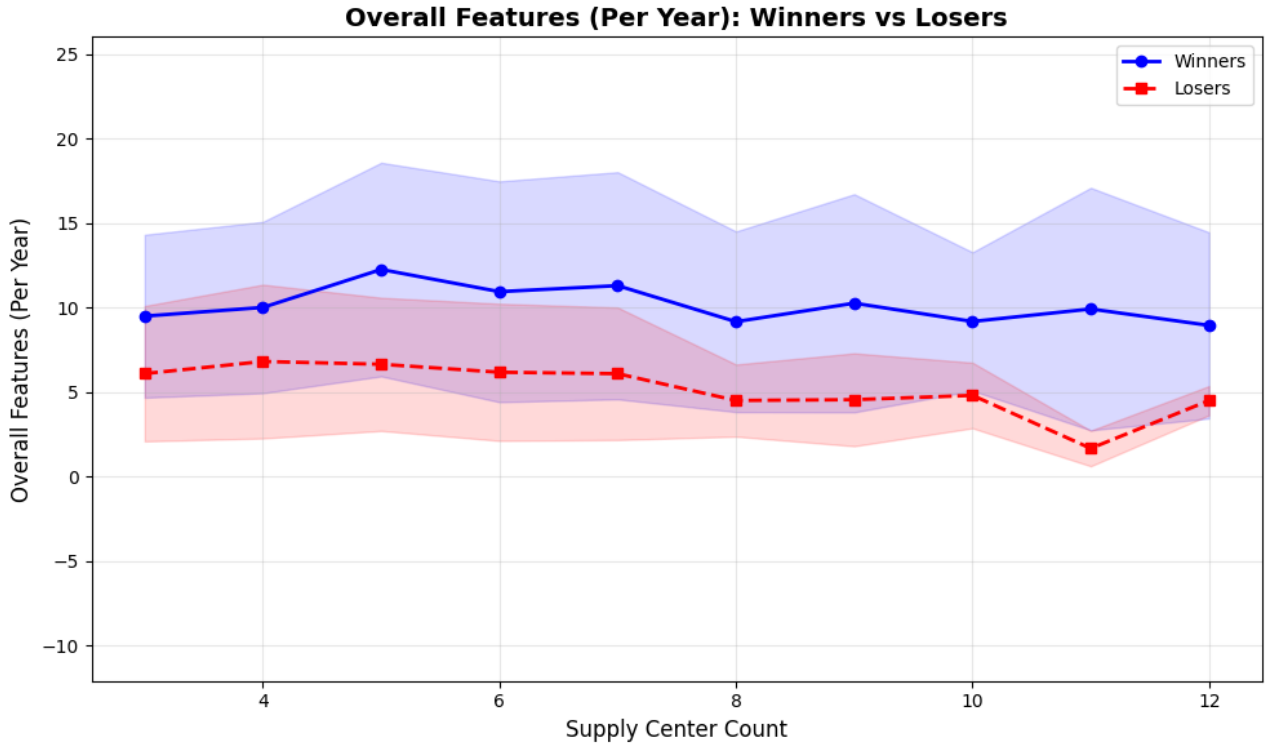}
    \caption{Number of negotiation tactics per year across supply center count}
    \label{fig:overall_feature_center_count}
\end{figure}

\Cref{fig:overall_feature_center_count} shows the changes in the overall prevalence as a player's supply-center count grows, which implicitly reflects the progression from the early to the late stages of the game. This figure aggregates all eight negotiation strategies into a single curve, highlighting the positional strengths of each strategy as a global trend. Notably, the observed trend underscores the importance of consistently employing negotiation tactics: throughout every phase of the games, winners exhibit a higher frequency of negotiation tactics compared to losers. This persistent difference demonstrates that winners' eventual successes are tied to sustained negotiation activity. The corresponding statistical significance tests are provided in \Cref{appn:long_term_significance_test}.

\section{LLM and Human Negotiation Tactics}

\subsection{LLM and Humans have Different Negotiation Tactics}
Our preceding analysis on the human WebDiplomacy corpus demonstrates that game success is indeed associated with the negotiation tactics. Building on this foundation, one core aim of this work is to probe the capabilities of LLMs as negotiators within the Diplomacy setting: How closely do LLM negotiators approximate human negotiation tactics, and can their negotiation style be steered to better align with high-quality human behavior?
We systematically examine whether LLMs exhibit substantive gaps compared to humans in the use of these tactics, and whether alignment can bridge this gap. To this end, we utilize LLMs to participate in negotiations and assess their negotiation proficiency. Leveraging an adapted version of the \textsc{sotopia} \citep{zhou2024sotopiainteractiveevaluationsocial} evaluation framework, we conducted one-on-one negotiation experiments between LLMs.
Each experiment focuses on a single one-on-one exchange, isolating each model’s style under realistic conversational pressure.


From the WebDiplomacy human-gameplay corpus, we sample 1{,}000 negotiation phases. Each phase comprises all messages exchanged immediately before players committed their orders, and the subsampled phases span diverse points in the game timeline (early, mid, or late game). We assign an LLM-negotiator agent the role of one player and prompted it to craft a reply to its partner's last message, negotiating game orders based on the current game board. The prompt instructs the model to balance tactical short-term gains (e.g., securing support for an attack) with relationship-building long-term goals (e.g., cultivating alliances), thereby mirroring the dual-goal orientation of skilled human players \citep{jaidka2023takesnegotiatemodelingsocial}. LLM-negotiators were constructed with the following models: Llama3.1‑8B‑Instruct \citep{dubey2024llama}, R1-distilled-LLama3-8B \citep{deepseekai2025deepseekr1incentivizingreasoningcapability}, and Magistral-2506-24B \citep{rastogi2025magistral}. The full prompt is in \Cref{appendix:prompt_templates}. 

\begin{figure}[ht]
    \centering
    \includegraphics[width=1\columnwidth]{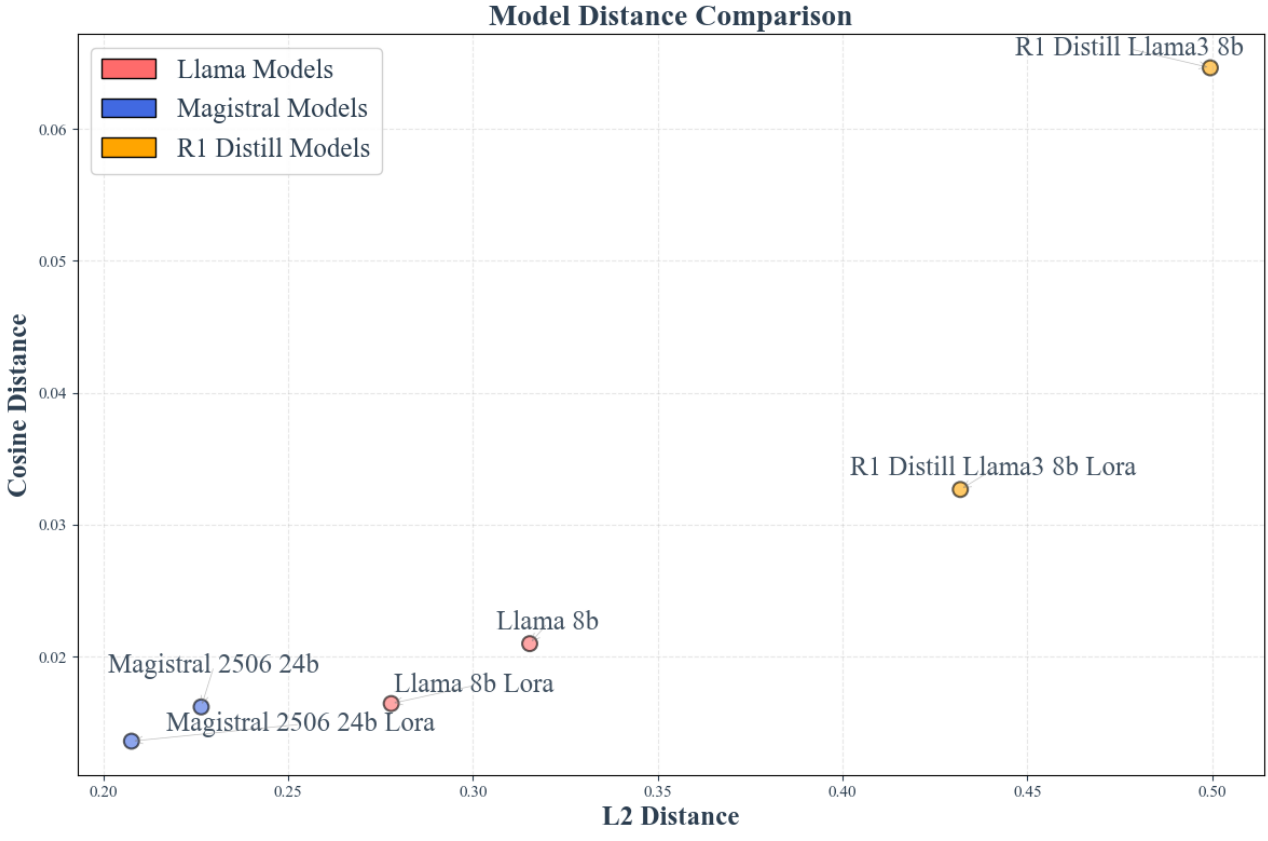}
    \caption{Model-Human Distance}
    \label{fig:overall_human_model_distance}
\end{figure}

We score each message produced by the LLM-negotiator with our LLM-as-a-judge pipeline (see Section~\ref{sec:annotations}). For each of the negotiation dimensions present, we recorded (i) the raw count of occurrences and (ii) a length-normalized rate of negotiation dimensions per sentence. These features are then aggregated into stylistic profiles for each model, enabling insight into whether the LLM style favors strategic maneuvers (e.g. game-move and information share) or social tactics (e.g. rapport and apologies).


To quantify the difference in negotiation techniques between humans and LLMs, we define $\mathcal{P}$ as the set of phases that contained both human and model utterances. For a phase $p\!\in\!\mathcal{P}$ and speaker $s$
(human or LLM), the LLM-judge with mturk instruct emits a binary value $\mathbf{f}_{s,p}\!\in\!\{0,1\}^{8}$, which is normalized by sentence count, $\tilde{\mathbf{f}}_{s,p}=\mathbf{f}_{s,p}/\text{sent\_cnt}(s,p)$.
Averaging over phases yields an 8-D \emph{mean style vector} (see \Cref{tab:negotiation_tactics}) per speaker:
\begin{equation}
\label{eq:averages}
\mathbf m_k = \frac{1}{|\mathcal P|}
              \sum_{p\in\mathcal P}\tilde{\mathbf f}_{k,p},
\qquad
\mathbf h   = \frac{1}{|\mathcal P|}
              \sum_{p\in\mathcal P}\tilde{\mathbf f}_{\text{hum},p}.
\end{equation}


We quantify LLM–human divergence with three distances, where lower values mean closer LLM-human alignment. We drew $1,000$ bootstrap samples over $\mathcal{P}$ and computed the metrics (see more results in \Cref{appn:examples_llm_vs_human}).
\begin{align}
\label{eq:distance_metrics}
\text{L2}_k      &= \lVert \mathbf m_k - \mathbf h \rVert_{2} \\[2pt]
\text{CosDist}_k &= 1 -
  \frac{\mathbf m_k^{\!\top}\mathbf h}
       {\lVert \mathbf m_k \rVert_{2}\,\lVert \mathbf h \rVert_{2}} 
\end{align}

\Cref{fig:all_distance_metrics,fig:overall_human_model_distance} reveals substantial gaps between human negotiation tactics and current LLM models. Magistral-2506-24B achieved the closest LLM-human distance. 
In contrast, the R1-distilled-Llama3-8B model showed a marked divergence from human reference. This suggests that the math-and-coding reasoning focus of this model might have a limited ability to mimic human negotiation tactics in our domain, underscoring the necessity of aligning reasoning models for social abilities (e.g., negotiation in our study) and broader interaction competence.

A feature-level analysis (\Cref{fig:features_vs_l2_distance}) further demonstrates that these gaps are not uniform across negotiation tactics. Specifically, when comparing LLMs, the R1-distilled-Llama3-8B model exhibits the largest deviations from human behavior on key tactical features (e.g. \textsc{Game-Move} and \textsc{Share-Information}), underscoring a pronounced deficit in emulating human-like strategic maneuvers. Meanwhile, the R1-distilled and Llama3.1-8B models display more modest, yet persistent, discrepancies on social-emotional and interpersonal features. All models, however, consistently underperform on subtle behaviors like \textsc{Personal-Thoughts} and \textsc{Reassurance}, indicating a general limitation in capturing the nuanced, relational aspects of human negotiation. 

Taken together, relying exclusively on reasoning-oriented distillation not only fails to align models with human style but may in fact exacerbate this misalignment. This limitation becomes especially salient in social reasoning tasks. These insights motivate the necessity of incorporating social reasoning and human-grounded data in future alignment efforts, which therefore motivates our subsequent style-alignment experiments. We show some examples in the \Cref{appn:examples_llm_vs_human}.




\subsection{Aligning LLM-Negotiation Tactics with Human}
In this section, we evaluate whether fine-tuning LLMs on human negotiation data produces negotiation tactics closer to human tactics.

The regression analysis performed in Section~\ref{sec:shorttermsuccess} shows that higher-order social tactics were predictors of subsequent growth. The full negotiation style distribution is shown in \Cref{fig:train_data_feature_frequency}. We focus on successful human dialogue, as measured by phases with $\Delta\mathrm{SC}$ increasing. 
Filtering the WebDiplomacy corpus for such phases yields 18{,}420 dialogue turns. We used these turns as a supervision corpus of effective human negotiation tactics. We use Supervised Fine-Tuning (SFT) on each model to steer the LLMs towards a more human-like distribution of negotiation strategies. Details are in \Cref{appendix:prompt_templates}.


Our quantitative analysis (see \Cref{fig:overall_human_model_distance} , \Cref{fig:all_distance_metrics} and \Cref{fig:lora_vs_base_distance}) 
demonstrates that instruction fine-tuning on the human-grounded negotiation corpus narrowed the gap between LLM-generated and human negotiation tactics. Specifically, \Cref{fig:average_lora_effect_across_model,fig:average_lora_effect_across_features} reveals that across most social negotiation features, all fine-tuned models exhibited reduced LLM-Human L2 distances. There were particularly strong convergence on social features of \textsc{Rapport}, \textsc{Compliment}, and \textsc{Apologies}. However, features like \textsc{Personal-Thoughts} and \textsc{Reassurance} remained more challenging, showing persistent LLM-human gaps. 

The difference plots further show that LoRA-based SFT produced the most pronounced distance reductions for the most different R1-Distill-Llama3-8B model, indicating substantial stylistic shift toward (Cosine=2.2\%) human-like negotiation. For models that already exhibited strong human alignment, such as Mistral-2506-24B, LoRA fine-tuning still yielded additional improvements (Cosine = 0.7\%), further aligning to human distributions. These results confirm that SFT with LoRA effectively enhances the alignment of LLM and human negotiation tactics.


Overall, fine-tuned models acquired more human-aligned behaviors that led to consistent reductions in differences between LLM and humans across all eight negotiation tactics (see \Cref{appn:examples_llm_negotiators_in_different_style}, \Cref{tab:negotiation_samples} and \Cref{fig:all_distance_metrics}). This convergence helps to validate the eight negotiation tactics as reliable proxies for human-grounded negotiation tactics and their utility as measurement tools and optimization targets, reinforcing their value as meaningful descriptors and effective behavioral targets.


\section{Conclusion}
We developed a reliable LLM-as-a-judge pipeline to annotate the WebDiplomacy dataset comprising of 4000 human-human Diplomacy game for a taxonomy of tactics based on Aristotle's Ethos-Pathos-Logos framework. These tactics are predictive of both short-term turn-to-turn success and long-term end-game success. The most predictive features are: \textsc{game move}, socio-emotional cues (\textsc{rapport}, and \textsc{reassurance}). We then prompted LLMs to reply to last turn messages as negotiators. While LLMs start off by having different negotiation tactics than humans, Supervised Fine-Tuning techniques can shift the tactics LLMs used to align better with the tactics real humans use. Our results lay the foundation for the effectiveness of online negotiation strategies, providing directions towards measuring the ability of LLM-agents to use negotiation tactics in a human-like way.


\section{Limitations}



\paragraph{Lack of Direct Game-Play Evaluation}
Our analysis focuses exclusively on negotiation tactics and the alignment between human and LLM behaviors, and does not evaluate whether fine-tuned LLM agents actually achieve higher win rates or improved game outcomes when deployed as autonomous Diplomacy players. Direct assessment of win rates and strategic success is out of the scope of this work. Instead, we aim to provide a detailed understanding of negotiation tactics and their human–LLM alignment. Future work should integrate end-to-end evaluations, placing aligned LLM agents into live game environments to determine whether improved tactic alignment ultimately translates into concrete strategic gains.

\paragraph{Potential Biases in Human and Model Data}
Our approach prioritizes learning and aligning with human negotiation tactics, but does not systematically filter or analyze for undesirable content such as social biases, toxicity, or hate speech that may be present in human data and potentially learned by LLMs during fine-tuning. As a result, the models may inherit and propagate problematic patterns observed in the training corpus. Further research should include dedicated analyses for bias and toxicity, and the development of mitigation strategies to ensure that aligned negotiation agents remain ethical and fair in their interactions.

\section*{Acknowledgments}
We thank WebDiplomacy for supporting this research by providing access to online gameplay data.
We are grateful to Mona Diab and Maarten Sap for helpful feedback on earlier versions of this work.
This material is based upon work supported by the Defense Advanced Research Projects Agency (DARPA) under Agreement No.
HR00112490410. 





\bibliography{custom}

\appendix

\section{Diplomacy Game Setting}
\label{appn:diplomacy_description}

Diplomacy is a strategic multi-agent negotiation game originally designed for seven players, each representing a major European power (Austria-Hungary, England, France, Germany, Italy, Russia, and Turkey). The game is played in discrete rounds corresponding to historical seasons (Spring and Fall), with each round consisting of two key phases: a negotiation phase and an order submission phase. During the negotiation phase, players may exchange private messages with any other player, formulating alliances, making promises, or attempting deception. The number of exchanged messages varies depending on the level of engagement, but in experimental settings, a typical round yields dozens of bilateral messages per player. After negotiations, all players simultaneously submit their movement orders for military units, which are then resolved according to deterministic rules of conflict resolution. A single round (negotiation plus order resolution) usually lasts from several minutes to an hour in controlled experimental contexts, though in traditional play by mail or online platforms, a round can span 24 to 72 hours.

Victory conditions in Diplomacy are defined by territorial control. The game board consists of 75 provinces, of which 34 contain supply centers. Each power begins with three or four supply centers, and control over a center determines the number of units a player may sustain. Players gain or lose units depending on the number of centers they control after each Fall phase. The ultimate objective is to capture at least 18 supply centers, which constitutes an outright win. Alternatively, when no single player can achieve this threshold, the game may end in a draw among the surviving players. This combination of simultaneous action resolution, unmediated negotiation, and long-term strategic planning makes Diplomacy a canonical testbed for studying cooperation, competition, alliance formation, and deception in multi-agent interaction settings.

\section{Definitions and Sociological Grounding for Negotiation Tactics}
\Cref{tab:negotiation_tactics_sociological} presents the eight negotiation tactics used in our taxonomy, their definitions (which are adapted from \citep{jaidka2023takesnegotiatemodelingsocial}), and the sociological grounding of each tactic.

\begin{table*}[ht]
\centering
{\small
\begin{tabularx}{\linewidth}{l l Y Y}
\toprule
Rhetoric & Negotiation Tactic & Definition & Sociological Grounding \\
\midrule
Ethos & Game-Move & Messages related to plans, thoughts and goals about a Diplomacy move & Game theory's emphasis that decisions of cooperate or compete are based on analysis of possible outcomes \cite{kibris2010cooperative} \\
Ethos & Share-Information & Messages about the history of or information gained about another player's move (except the speaker's and recipient's) & Quantity \& quality of information shared is typically associated with negotiation effectiveness \cite{butler1999trust} \\
Logos & Reasoning & Speculative reasoning, justification of past or future moves & Receivers are more likely to agree with speakers who provide strong factual evidence and rational arguments \cite{BRETT201668} \\
Pathos & Rapport & Messages that build trust and mutual understanding between speaker and received & Building rapport builds trust \cite{kim2015emotional}, and higher trust increases negotiation success \cite{lewicki2013role} \\
Pathos & Apologies & Expressions of regrets or remorse about past moves & Repairs both competence-based and integrity-based trust violations \cite{lewicki2013trust} \\
Pathos & Personal-Thoughts & Messages that reflect the speaker's inner reflections, opinions or feelings & Build trust by demonstrating vulnerability with self-disclosure \cite{altman1973social} \\
Pathos & Compliment & Positive messages about the recipient or recipient's moves & Emotional regulation strategy to enhance trust \cite{kim2015emotional} \\
Pathos & Reassurance & Supportive messages to restore confidence in recipient's game position & Emotional regulation strategy to enhance trust \cite{kim2015emotional} \\
\bottomrule
\end{tabularx}
}
\caption{Taxonomy of Negotiation Tactics, Definitions (adapted from \citet{jaidka2023takesnegotiatemodelingsocial}), and sociological grounding}
\label{tab:negotiation_tactics_sociological}
\end{table*}

\section{Datasets Information}
\label{appn:datasets_details}
This study used two datasets: (1) It Takes Two and (2) WebDiplomacy. Here are the details of these two datasets. 

The It Takes Two dataset was originally collected by~\cite{peskov2020takes}, which contains messages passed during the Diplomacy game from recruited human players. This dataset was further processed by~\cite{jaidka2023takesnegotiatemodelingsocial} filtered for meaningful messages that contained more than five words and annotated for fine-grained negotiation tactics.
This dataset contains 11,366 messages from 10 games. 

The WebDiplomacy dataset is licensed from the server administrator of the WebDiplomacy platform (\url{https://www.webdiplomacy.net}), an online platform to play the Diplomacy game. This admin first filtered the WebDiplomacy games for games with messages. Next he applied a player filter. He selected for quality players: players with more than 5 games joined, an ELO rating over 105, points 120, reliability over 70, had more than one game won, and more than 5\% game-win ratio. Then, he used the set of filtered players to select regular-press games that had more than 3 of these quality players in the game. 
From those games, the admin excluded those with No Moves Received (NMRs), and sampled 4000 games. 
This dataset contains 4000 games with the following information (see \Cref{tab:webdip_schema})

\section{LLM-as-a-Judge Template}
\label{appn:llm_as_a_judge_template}

\begin{figure*}[htbp]
    \centering
    \includegraphics[width=0.95\textwidth]{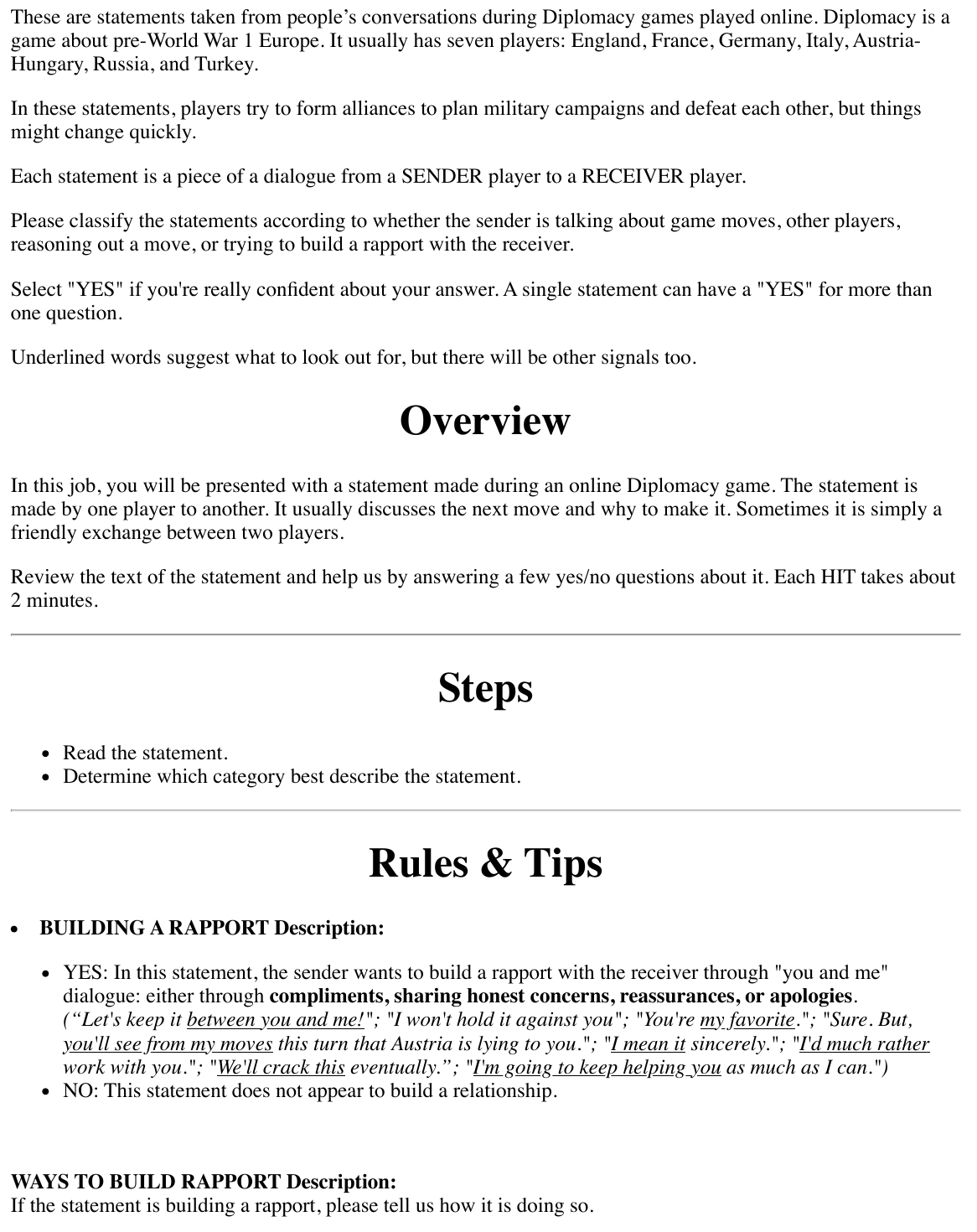}
    \caption{Instruction as MTurk for expert annotators (page 1)}
    \label{fig:mturk_instruct_1}
\end{figure*}

\begin{figure*}[htbp]
    \centering
    \includegraphics[width=0.95\textwidth]{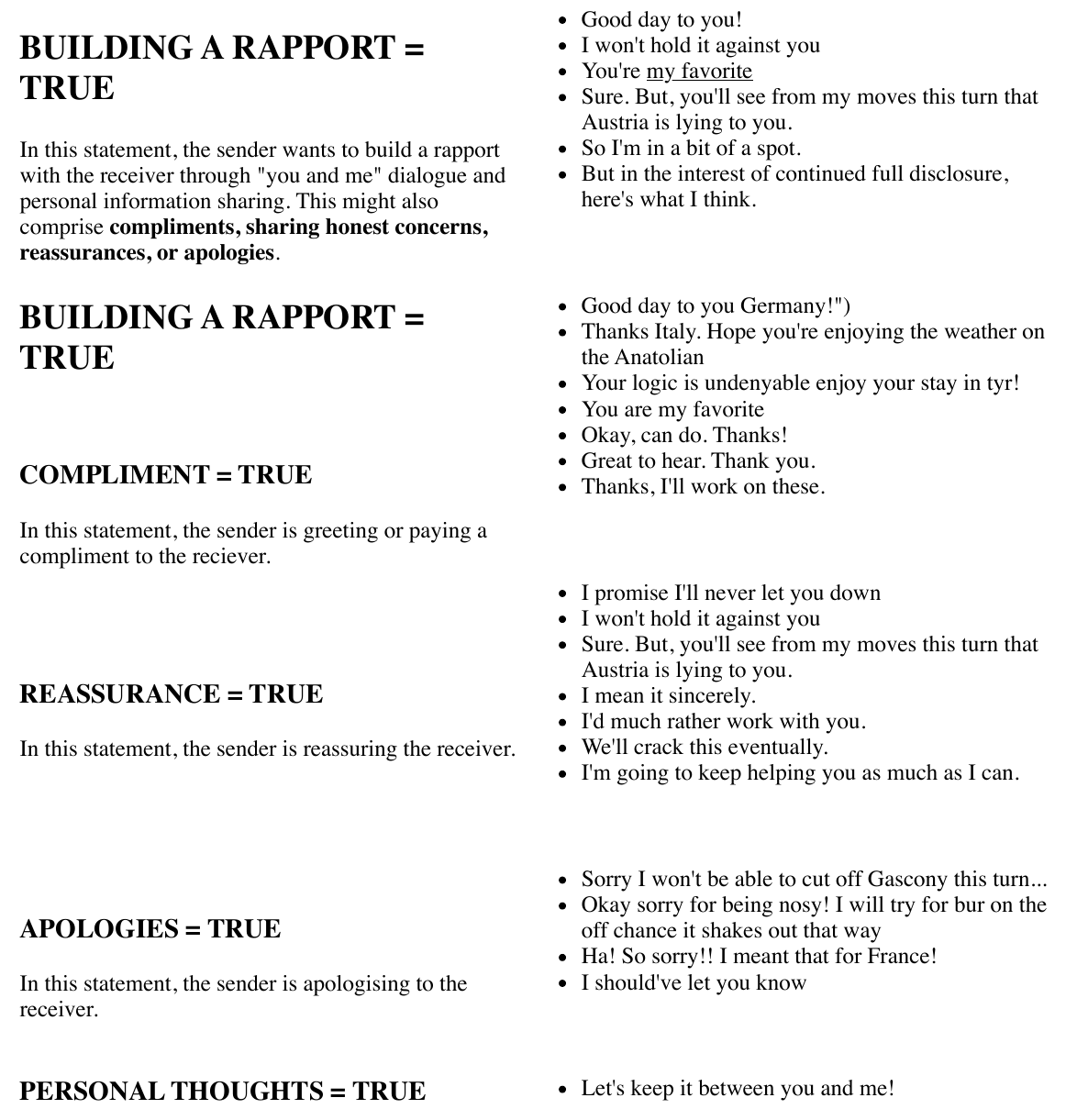}
    \caption{Instruction as MTurk for expert annotators (page 2)}
    \label{fig:mturk_instruct_2}
\end{figure*}

\begin{figure*}[htbp]
    \centering
    \includegraphics[width=0.95\textwidth]{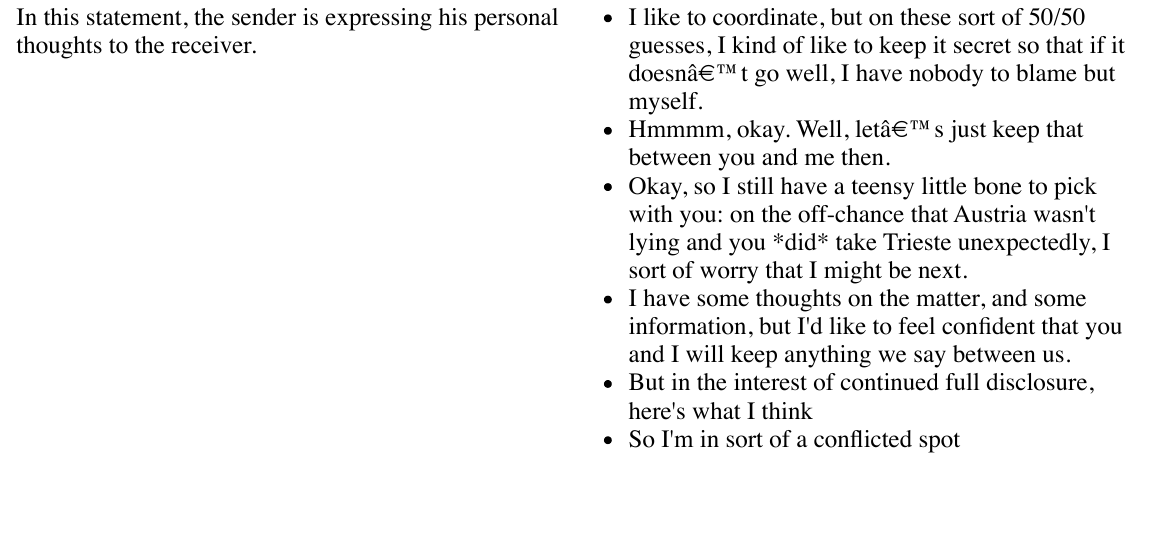}
    \caption{Instruction as MTurk for expert annotators (page 3)}
    \label{fig:mturk_instruct_3}
\end{figure*}

In the LLM-as-a-Judge setup, the baseline (zero-shot) instruction template is presented in \Cref{tab:baseline_judge_prompt_full}, while the few-shot variant, which gives exampless under each question, is shown in \Cref{tab:fewshot_judge_prompt_full}. The MTurk instruction (without giving the examples) and the MTurk-with-examples variant are illustrated in \Cref{fig:mturk_instruct_1,fig:mturk_instruct_2,fig:mturk_instruct_3}.

\section{Detailed Discussion of LLM-as-a-Judge Agreement with Human}
\label{appn:llm_as_a_judge_details}

\begin{figure*}[htbp]
    \centering
    \includegraphics[width=0.95\textwidth]{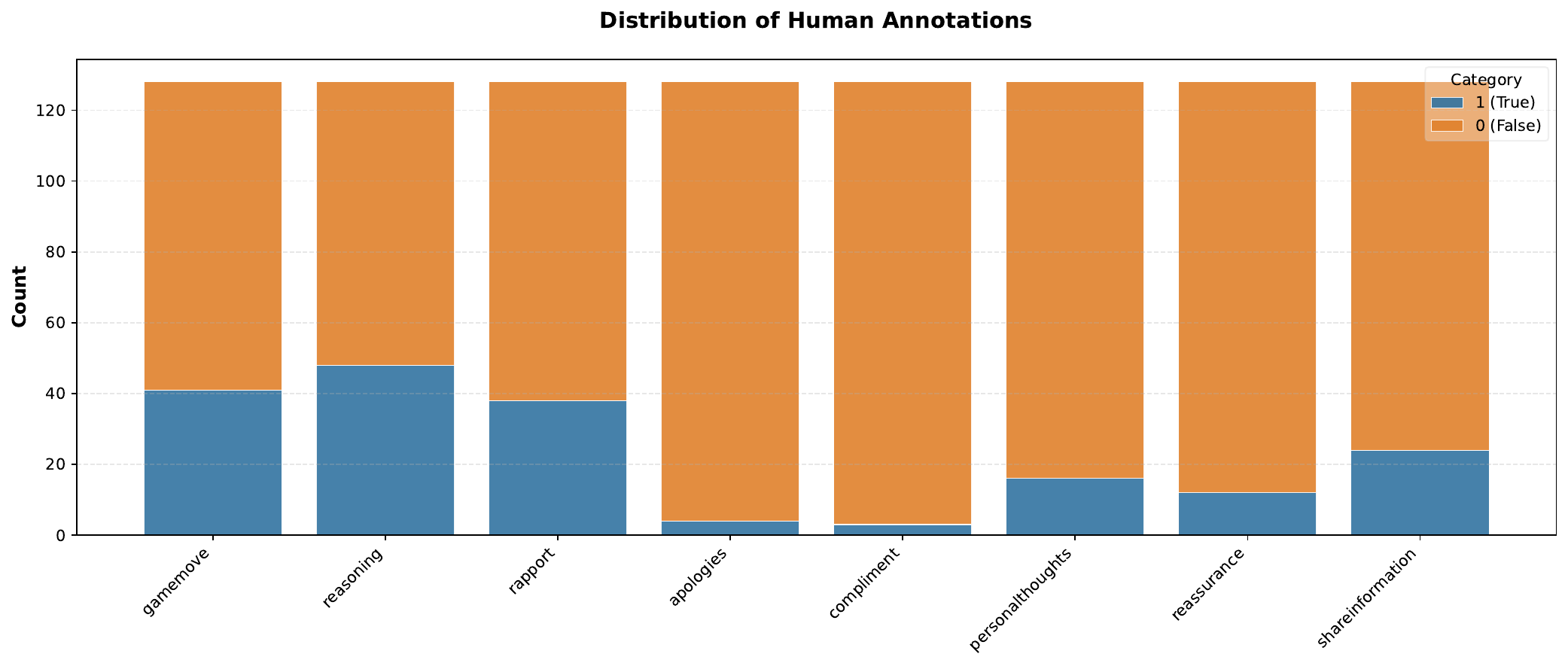}
    \caption{The Distribution of Human Annotation}
    \label{fig:human_annotation_distribution}
\end{figure*}

\paragraph{Why Gwet’s AC1 (instead of Fleiss’ $\kappa$).}
Our annotation labels are highly imbalanced across categories (see \Cref{fig:human_annotation_distribution}), with several tactics having very low “True’’ prevalence (e.g., \textit{apologies}, \textit{compliment}, \textit{personal thoughts}). In such settings, Fleiss’ $\kappa$ is known to suffer from the “$\kappa$ paradox’’: even when raters agree on most items, $\kappa$ can be deflated toward low values when the marginal distributions are skewed or when there is systematic class imbalance \citep{feinstein1990high,byrt1993bias,hallgren2012computing}. Intuitively, $\kappa$’s chance-agreement term $P_e$ is computed directly from the observed marginals; under high or low prevalence this inflates the expected agreement and depresses the coefficient, producing misleadingly “low’’ reliability.

Gwet’s AC1 addresses this issue by using a more stable estimator of chance agreement that corrects the bias induced by extreme or unequal marginals \citep{gwet2002kappa,gwet2008computing}. AC1 preserves the same interpretability as $\kappa$—$1$ indicates perfect agreement and $0$ indicates chance-level—but its chance-agreement component is far less sensitive to prevalence, yielding reliability estimates that better reflect actual rater concordance under class imbalance. Comparative studies consistently find that AC1 remains robust where $\kappa$ becomes paradoxically small in unbalanced, binary, or sparse multi-category settings \citep{wongpakaran2013comparison,gwet2014handbook}. Given the strong skew evident in our data, we therefore report Gwet’s AC1 as our primary agreement coefficient and include Fleiss’ $\kappa$ only for completeness. This choice avoids underestimating reliability due to prevalence effects and aligns with best practices for imbalanced annotation tasks.

\paragraph{Empirical evidence in our annotations.}

\begin{figure*}[htbp]
    \centering
    \includegraphics[width=1\textwidth]{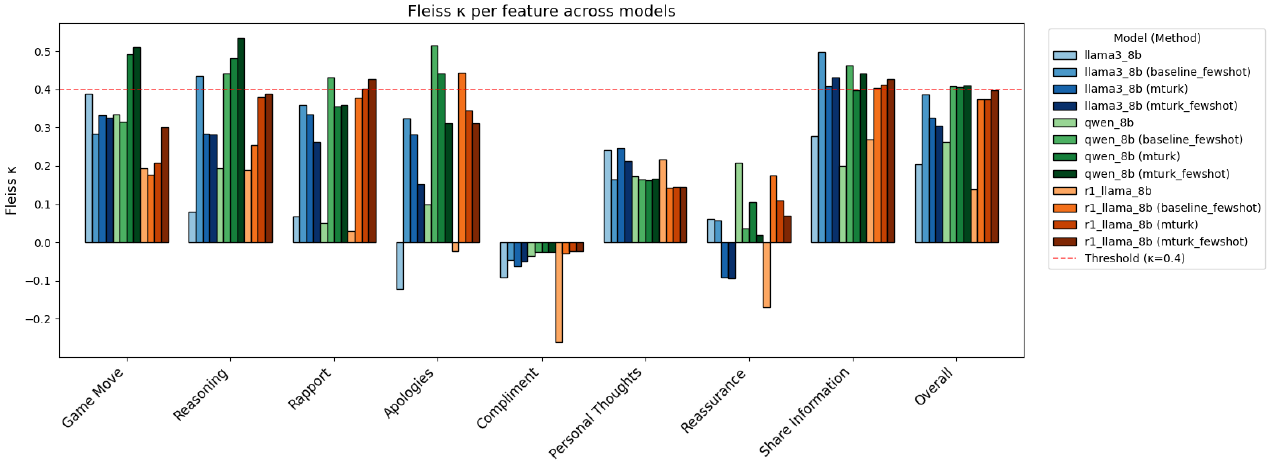}
    \caption{Fleiss’ $\kappa$ scores for human annotations agreement}
    \label{fig:fleiss_kappa_by_family}
\end{figure*}

\begin{figure*}[htbp]
    \centering
    \includegraphics[width=1\textwidth]{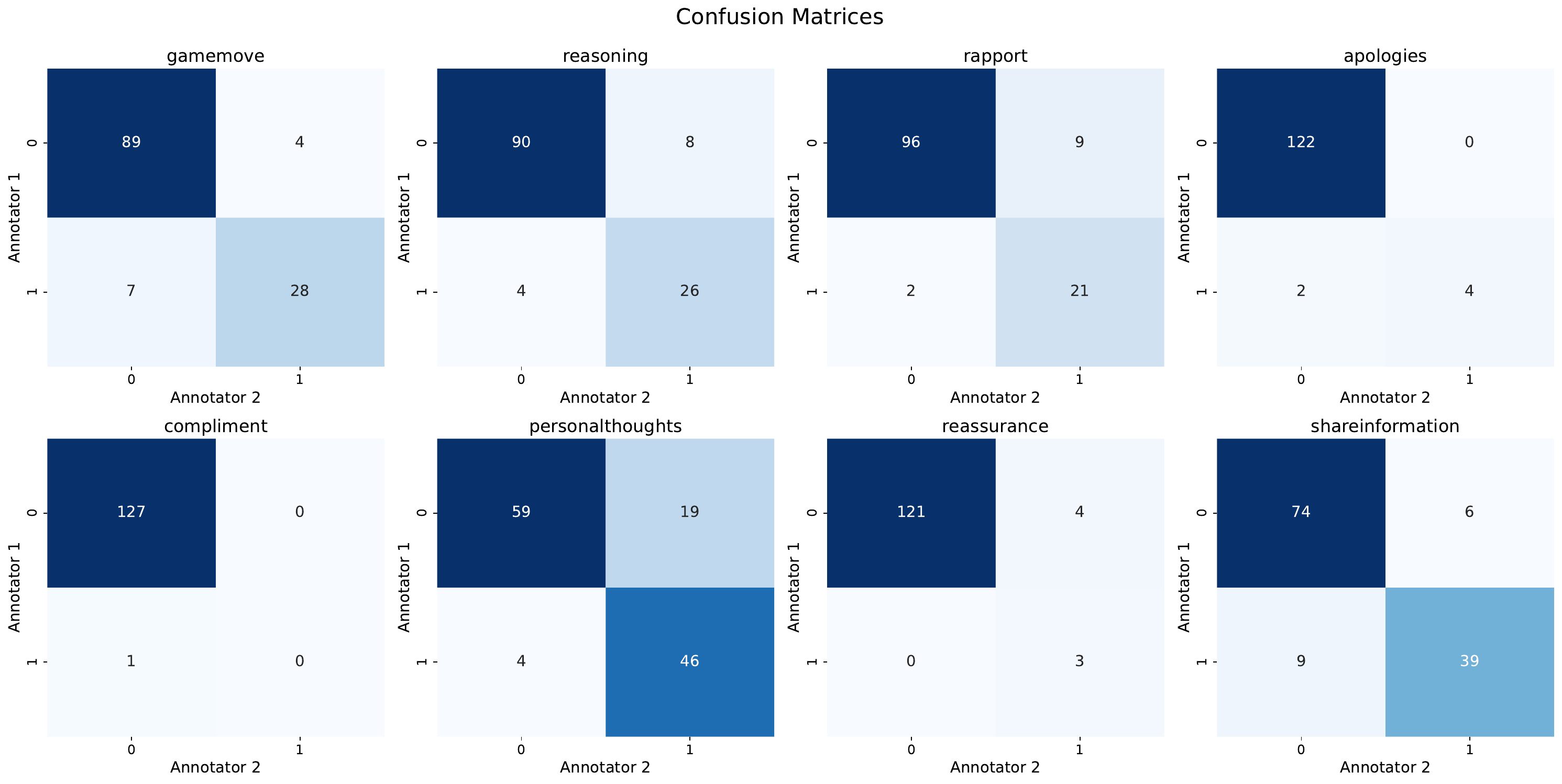}
    \caption{Confusion matrices for human annotations agreement}
    \label{fig:confusion_matrices}
\end{figure*}

The stacked counts in \Cref{fig:human_annotation_distribution} make the prevalence skew explicit, with “False’’ dominating most categories. Despite this imbalance, the human–human confusion matrices (see \Cref{fig:confusion_matrices}) are strongly diagonal, indicating high observed agreement. Consistent with the literature, AC1 yields substantively higher and crucially more faithful estimates of reliability than Fleiss’ $\kappa$ (see \Cref{fig:fleiss_kappa_by_family}) in these categories, reflecting that annotators largely agree even when positives are rare.

\section{Prompt Templates for LLM Negotiators}
\label{appendix:prompt_templates}

\begin{table*}[ht]
\centering
\small
\begin{tabularx}{\textwidth}{lX}
\toprule
\textbf{Element} & \textbf{Description} \\
\midrule
id & Unique identifier for the game \\
map & Game map type (e.g., standard) \\
rules & List of rules used in the game \\
phases & List of all game phases; each phase contains: \\
name & Name of the phase (e.g., S1901M) \\
state: & Game state for the phase, including: \\
timestamp & Timestamp for the phase \\
zobrist\_hash & Hash representing the board state \\
note & Miscellaneous notes on the phase \\
name & Name of the phase (redundant) \\
units: & Current unit positions for each power \\
<POWER> & List of units for each power (e.g., ['A BUD', ...]) \\
retreats: & Retreat status for each power \\
<POWER> & Retreat information for each power \\
centers: & Controlled supply centers for each power \\
<POWER> & List of supply centers for each power \\
homes: & Home centers for each power \\
<POWER> & List of home centers for each power \\
influence: & Regions influenced by each power \\
<POWER> & List of influenced regions for each power \\
civil\_disorder: & Civil disorder status for each power \\
<POWER> & 0 (normal) or 1 (civil disorder) \\
builds: & Build/disband information for each power \\
<POWER>: & Details for each power \\
count & Number of builds/disbands for each power \\
homes & Possible build locations for each power \\
game\_id & Game ID (redundant) \\
map & Map type (redundant) \\
rules & List of rules (redundant) \\
orders: & Player orders for the phase \\
<POWER> & List of orders submitted by each power \\
results: & Adjudication results for each unit/location \\
<UNIT/LOCATION> & Result list for the specified unit or location \\
messages: & List of all messages for the phase; each message contains: \\
sender & Sending player (power/country) \\
recipient & Recipient (power/country or GLOBAL for broadcast) \\
time\_sent & Time the message was sent \\
phase & Phase during which the message was sent \\
message & Content of the message \\
\bottomrule
\end{tabularx}
\caption{Structure of the WebDiplomacy dataset. Each game consists of multiple phases, with each phase recording the full board state, player orders, adjudication results, and negotiation messages.}
\label{tab:webdip_schema}
\end{table*}

\begin{table*}[t]
\centering
\begin{tabular}{p{\textwidth}}
\toprule
\textbf{Instruction for Utterance-level Strategy Classification} \\
\midrule
These are statements taken from people’s conversations during Diplomacy games played online. Diplomacy is a game about pre-World War 1 Europe. It usually has seven players: England, France, Germany, Italy, Austria-Hungary, Russia, and Turkey.

In these statements, players try to form alliances to plan military campaigns and defeat each other, but things might change quickly.

Each statement is a piece of a dialogue from a \textbf{SENDER} player to a \textbf{RECEIVER} player.

Please classify the statements according to whether the sender is talking about game moves, other players, reasoning out a move, or trying to build a rapport with the receiver.

Select \texttt{YES} if you're really confident about your answer. A single statement can have a \texttt{YES} for more than one question.

Underlined words suggest what to look out for, but there will be other signals too. \\

\addlinespace
\textbf{For each of the following questions, answer \texttt{YES} if you are confident about your answer. A single statement can have a \texttt{YES} for more than one question. Underlined words suggest what to look out for, but there will be other signals too.}
\\
\midrule
\textbf{1. Is this statement about the sender's or receiver's GAME MOVE?} \\
The sender states an actual or suggested game move by the sender or the receiver. It might also be in the form of an acceptance, a question, or a clarification. \\
\addlinespace
\textbf{2. Does this statement PROVIDE REASONS for the sender's or receiver's move?} \\
The sender offers justification or explanations for a move by themselves or by the receiver, guesses what moves might happen next, or discusses a move that already happened. \\
\addlinespace
\textbf{3. Does this statement involve BUILDING a RAPPORT?} \\
In this statement, the sender wants to build a rapport with the receiver through "you and me" dialogue and personal information sharing. \\
\addlinespace
\textbf{4. Is the sender greeting or paying a COMPLIMENT to the receiver?} \\
The sender is greeting or paying a compliment to the receiver. \\
\addlinespace
\textbf{5. Is the sender offering REASSURANCE to the receiver?} \\
The sender is reassuring the receiver. \\
\addlinespace
\textbf{6. Is the sender APOLOGISING to the receiver?} \\
The sender is apologising to the receiver. \\
\addlinespace
\textbf{7. Is the sender SHARING PERSONAL THOUGHTS or feelings with the receiver?} \\
The sender is sharing their personal thoughts or feelings with the receiver. \\
\addlinespace
\textbf{8. Does this statement SHARE INFORMATION about other players?} \\
This statement shares information related to other game players, NOT the sender or the receiver. \\
\addlinespace
\midrule
\textbf{Expected Output Format:}
\begin{verbatim}
1. YES
2. NO
...
8. YES
\end{verbatim}

\textbf{Here is the statement:} \texttt{\{STATEMENT\}}
\\
\bottomrule
\end{tabular}
\caption{Full instruction prompt used for LLM-as-a-Judge utterance-level strategy annotation.}
\label{tab:baseline_judge_prompt_full}
\end{table*}

\begin{table*}[t]
\centering
\begin{adjustbox}{max totalsize={\textwidth}{0.9\textheight},center}
\begin{tabular}{p{\textwidth}}
\toprule
\textbf{Instruction for Utterance-level Strategy Classification (Few-shot)} \\
\midrule
These are statements taken from people’s conversations during Diplomacy games played online. Diplomacy is a game about pre-World War 1 Europe. It usually has seven players: England, France, Germany, Italy, Austria-Hungary, Russia, and Turkey.

In these statements, players try to form alliances to plan military campaigns and defeat each other, but things might change quickly.

Each statement is a piece of a dialogue from a \textbf{SENDER} player to a \textbf{RECEIVER} player.

Please classify the statements according to whether the sender is talking about game moves, other players, reasoning out a move, or trying to build a rapport with the receiver.

Select \texttt{YES} if you're really confident about your answer. A single statement can have a \texttt{YES} for more than one question.

Underlined words suggest what to look out for, but there will be other signals too. \\

\addlinespace
\textbf{For each of the following questions, answer \texttt{YES} if you are confident about your answer. A single statement can have a \texttt{YES} for more than one question. Underlined words suggest what to look out for, but there will be other signals too.}
\\
\midrule

\textbf{1. Is this statement about the sender's or receiver's GAME MOVE?} \\
The sender states an actual or suggested game move by the sender or the receiver. It might also be in the form of an acceptance, a question, or a clarification. \\
\textit{\textcolor{blue}{Examples: \{exmaple\}, \{exmaple\}, \{exmaple\}}} \\
\addlinespace

\textbf{2. Does this statement PROVIDE REASONS for the sender's or receiver's move?} \\
The sender offers justification or explanations for a move by themselves or by the receiver, guesses what moves might happen next, or discusses a move that already happened. \\
\textit{\textcolor{blue}{Examples: \{exmaple\}, \{exmaple\}, \{exmaple\}}} \\
\addlinespace

\textbf{3. Does this statement involve BUILDING a RAPPORT?} \\
In this statement, the sender wants to build a rapport with the receiver through ``you and me'' dialogue and personal information sharing. \\
\textit{\textcolor{blue}{Examples: \{exmaple\}, \{exmaple\}, \{exmaple\}}} \\
\addlinespace

\textbf{4. Is the sender greeting or paying a COMPLIMENT to the receiver?} \\
The sender is greeting or paying a compliment to the receiver. \\
\textit{\textcolor{blue}{Examples: \{exmaple\}, \{exmaple\}, \{exmaple\}}} \\
\addlinespace

\textbf{5. Is the sender offering REASSURANCE to the receiver?} \\
The sender is reassuring the receiver. \\
\textit{\textcolor{blue}{Examples: \{exmaple\}, \{exmaple\}, \{exmaple\}}} \\
\addlinespace

\textbf{6. Is the sender APOLOGISING to the receiver?} \\
The sender is apologising to the receiver. \\
\textit{\textcolor{blue}{Examples: \{exmaple\}, \{exmaple\}, \{exmaple\}}} \\
\addlinespace

\textbf{7. Is the sender SHARING PERSONAL THOUGHTS or feelings with the receiver?} \\
The sender is sharing their personal thoughts or feelings with the receiver. \\
\textit{\textcolor{blue}{Examples: \{exmaple\}, \{exmaple\}, \{exmaple\}}} \\
\addlinespace

\textbf{8. Does this statement SHARE INFORMATION about other players?} \\
This statement shares information related to other game players, NOT the sender or the receiver. \\
\textit{\textcolor{blue}{Examples: \{exmaple\}, \{exmaple\}, \{exmaple\}}} \\
\addlinespace

\midrule
\textbf{Expected Output Format:}
\begin{verbatim}
1. YES
2. NO
...
8. YES
\end{verbatim}

\textbf{Here is the statement:} \texttt{\{STATEMENT\}}
\\
\bottomrule
\end{tabular}
\end{adjustbox}
\caption{Full few-shot instruction prompt used for LLM-as-a-Judge utterance-level strategy annotation. \emph{Few-shot placeholders (3 per question) are shown in blue as \{exmaple\}; examples originate from MTurk-collected interaction snippets.}}
\label{tab:fewshot_judge_prompt_full}
\end{table*}

We condition the model on the current phase, the dyadic dialogue context, the most recent executed orders, and a compact snapshot of the board state (centers and units), then assign the model a single speaking role for the turn. This follows the CICERO dialogue‑agent design that situates language generation in the game state and recent conversation, while instructing the agent to advance plans through cooperative negotiation. \citep{meta2022human}

The following is the full prompt template used for the LLM acting as negotiator:

\begin{tcolorbox}[colback=gray!6!white, colframe=black!50, title=LLM Negotiator Prompt Template, arc=2mm, boxrule=0.5pt, left=1mm, right=1mm, top=1mm, bottom=1mm, fontupper=\small]
\textbf{SYSTEM}: You are playing the diplomacy game. You will negotiate with the other player so that it plays moves beneficial to your board position, either this turn or in future turns.

\textbf{You are in Phase}: \textcolor{blue}{\texttt{\{PHASE\_NAME\}}}

\textbf{The dialogue are between the two countries}: \textcolor{orange}{\texttt{\{COUNTRY1\}}} \textbf{and} \textcolor{orange}{\texttt{\{COUNTRY2\}}}

\textbf{The previous turn dialogue history is:}\\
\textcolor{teal}{\texttt{\{DIALOGUE\_HISTORY\}}}

\textbf{The previous order history is:}\\
\textcolor{violet}{\texttt{\{ORDER\_HISTORY\}}}

\textbf{This is the information of the current game state:}\\
\textbf{Centers:} \textcolor{magenta}{\texttt{\{CENTER\_INFO\}}}\\
\textbf{Units:} \textcolor{magenta}{\texttt{\{UNIT\_INFO\}}}

\textbf{You are playing as \textcolor{orange}{\texttt{\{COUNTRY1\}}}}. You are playing the diplomacy game, you will negotiate with the other player so that it will play moves that are beneficial to your board position, either this turn or in future turns.
\end{tcolorbox}

\section{Experiment details}
\label{appn:experiment_details}

\paragraph{LoRA Fine-tuning}
We performed alignment training using the Supervised Fine-Tuning (SFT) methods on LLaMA-3.1-8B-Instruct, Magistral-2506-24B, and R1-distilled-LLama3-8B. Both training approaches utilized the Low-Rank Adaptation (LoRA) technique \citep{hu2021loralowrankadaptationlarge}, which enabled efficient fine-tuning of the large language model by adapting a subset of its parameters. The experiments were conducted using 4 NVIDIA A6000 GPUs, with each GPU processing a batch size of 4.

For LoRA, we applied the technique across all layers of the model for SFT. The training configuration included a learning rate of $1.0\times 10^{-5}$, regulated by a cosine scheduler, a warm-up phase consisting of 100 steps, and a gradient accumulation over 8 steps. We didn't limit training to three epochs with a maximum sequence length. Each training required approximately 20-24 hours to complete. To optimize computational resources, we used mixed-precision training with bfloat16. Both datasets were preprocessed using each model family's template and split into training and validation sets, with 10\% of the data reserved for validation to monitor performance.

The training prompt for SFT follows the template below:

\begin{itemize}
  \item \textbf{Instruction}: You are playing diplomacy game, you will negotiate with the other player so that it will play moves that are beneficial to your board position, either this turn or in future turns.
  
  \item \textbf{Input (sender messages)}: England has told me that he will support his army into Belgium. I am happy to be allies with you against him, but I'd like Sweden. It seems to our mutual advantage for you to cut his support in the North Sea and attempt to bounce Belgium.

  \item \textbf{Output (recipient messages)}: I like the DMZ, but we'll have to see about Sweden; it depends on the actions of England and France, sorry.
\end{itemize}

\section{Examples of LLM Negotiators in Different Styles}
\label{appn:examples_llm_negotiators_in_different_style}
The picked examples in \Cref{tab:negotiation_samples} illustrate how LLMs base model and their LoRA‑tuned variants, occupy different negotiation tactics space. We annotate each utterance with eight features that span task‑oriented planning (Game Move, Reasoning, Share Information) and socio‑emotional maintenance (Rapport, Apologies, Compliment, Reassurance, Personal Thoughts). These qualitative snapshots complement our quantitative analyses by showing how specific stylistic cues surface in model outputs.

\begin{table*}[t]
\centering
\setlength{\tabcolsep}{4pt}   
\renewcommand{\arraystretch}{1.15}
\small
\begin{tabularx}{\textwidth}{lY*{8}{C}}
\toprule
\small\textbf{Model} & \textbf{Messages} &
\textbf{GM} & \textbf{Rea.} & \textbf{Rap.} & \textbf{Apo.} &
\textbf{Comp.} & \textbf{PT} & \textbf{Reass.} & \textbf{SI} \\
\midrule
Llama-8b & I'm considering moving my fleet from the Ionian to the AEG to counter Italy's fleet. What are your thoughts on this? & \tick & \tick & \cross & \cross & \cross & \cross & \cross & \tick \\
Llama-8b-Lora & I hope this game is going well for you so far. I must say, I'm a bit concerned about the situation in the Balkans, but I'm willing to work with you to find a solution that benefits both of our nations. & \cross & \cross & \tick & \tick & \cross & \cross & \tick & \cross \\
Magistral-2506-24b & I appreciate the offer, but I think I'd prefer to focus on consolidating our position first. Let's see how the board develops after the next few moves before we make any further commitments. I suggest we hold our current positions and observe Russia's next move, as their actions will likely dictate how we proceed. & \tick & \tick & \cross & \cross & \cross & \cross & \cross & \tick \\
Magistral-2506-24b-Lora & Understood. I'll coordinate with you to secure Liverpool next phase while you focus on the North Sea. Let's finalize our move orders to ensure a smooth execution. Also, I'll monitor Italy's movements to adjust our strategy if necessary. & \tick & \tick & \cross & \cross & \cross & \cross & \cross & \tick \\
R1-Distill-Llama3-8b & I appreciate our growing naval presence and the strength it brings to our alliance. Let's continue to work together to secure our shared borders and counter any threats. I'm confident that with our combined efforts, we'll achieve our goals. & \cross & \cross & \tick & \tick & \cross & \cross & \cross & \cross \\
R1-Distill-Llama3-8b-Lora & Thank you for your confidence in our strategy. I’m as committed as you are to this alliance, and I believe our combined strength will make us a formidable force against our common enemies. Let’s discuss a specific action we can take to solidify our position and strengthen our alliance further. & \cross & \cross & \tick & \tick & \tick & \cross & \tick & \cross \\
\bottomrule
\end{tabularx}
\caption{Sample negotiations from different models showing various negotiation features (\tick\ indicates the presence of a feature, while \cross\ indicates its absence). Abbreviations: GM = Game Move; Rea. = Reasoning; Rap. = Rapport; Apo. = Apologies; Comp. = Compliment; PT = Personal Thoughts; Reass. = Reassurance; SI = Share Information.}

\label{tab:negotiation_samples}
\end{table*}

\section{Examples of LLM Negotiation Different Style Different to Human}
\label{appn:examples_llm_vs_human}
\begin{figure*}[htbp]
    \centering
    \includegraphics[width=0.95\textwidth]{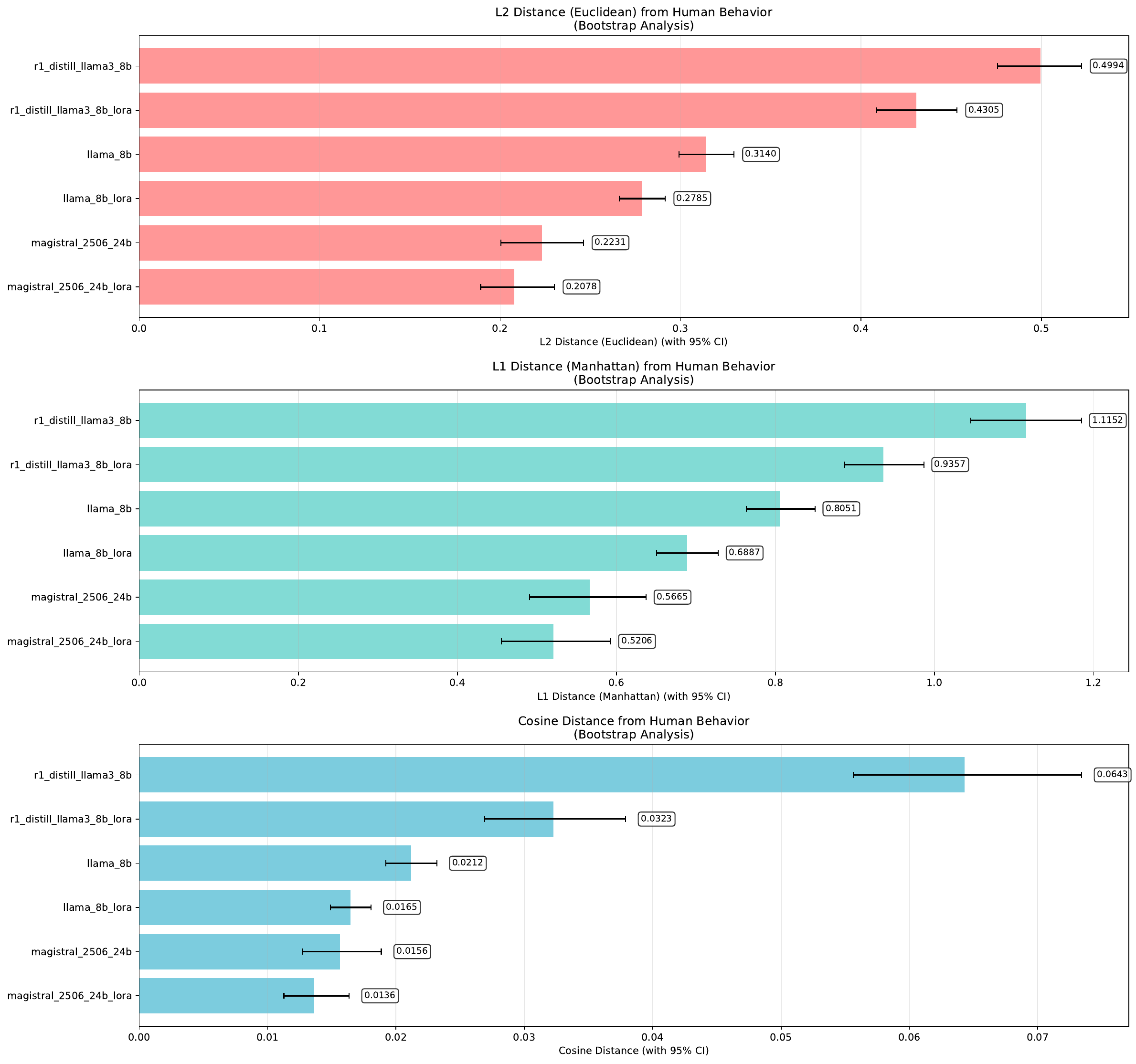}
    \caption{Model L2 distance from humans (lower is better).}
    \label{fig:all_distance_metrics}
\end{figure*}

To complement our quantitative analysis of LLM-human style divergence (\Cref{fig:all_distance_metrics,fig:overall_human_model_distance}), we present concrete examples and qualitative comparisons illustrating the nature of these differences. \Cref{tab:r1_pair,tab:llama3_pair,tab:mag_pair}, juxtapose negotiation utterances generated by different LLM models (with and without LoRA alignment) against randomly sampled human messages, revealing both the most and least human-like traits expressed by each model family.

\begin{figure*}[h]
    \centering
    \includegraphics[width=0.9\textwidth]{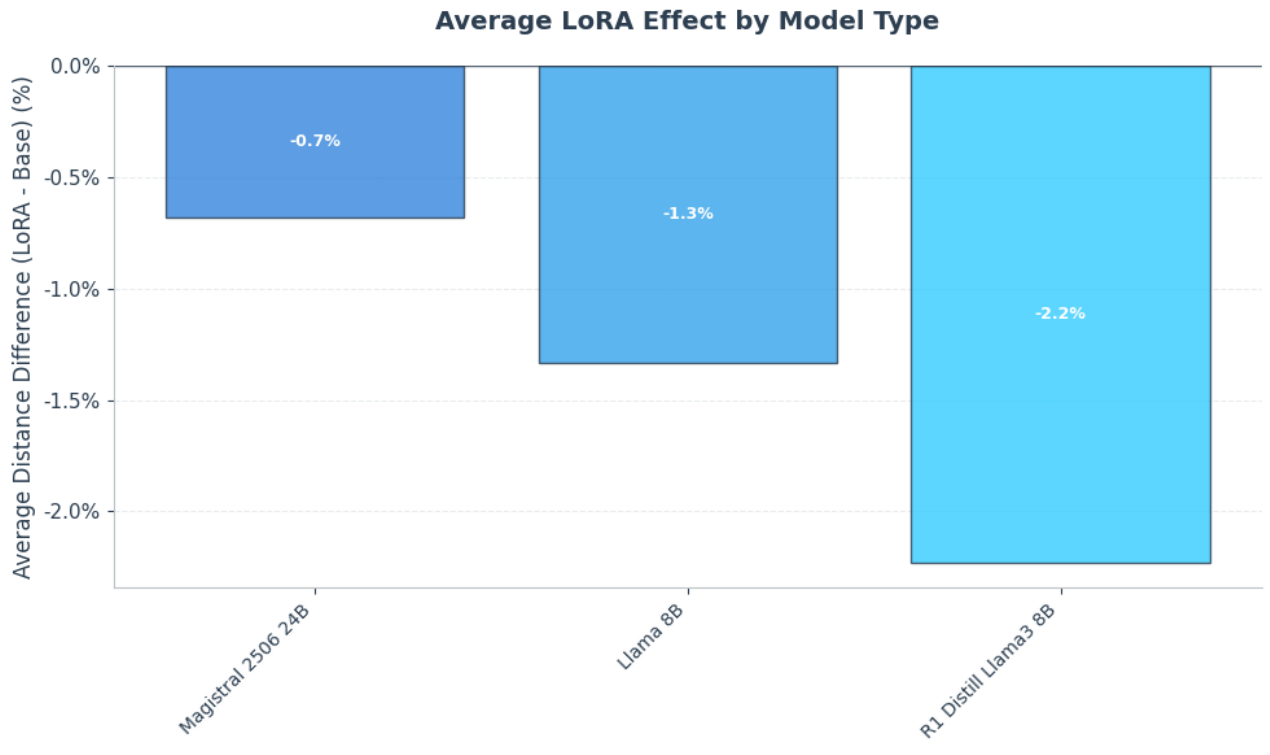}
    \caption{Average LoRA effect by model type. Bars show the mean change in L2 distance to the human reference (LoRA $-$ Base, in percentage points) averaged across the eight negotiation features; negative values indicate improvement (smaller distance).}
    \label{fig:average_lora_effect_across_model}
\end{figure*}

\paragraph{Deficits in Rapport and Socio-Emotional Expression.}
A clear and recurring shortfall across LLM outputs is their limited use of rapport-building language and socio-emotional cues. While LoRA-aligned variants show some improvement (see \Cref{fig:average_lora_effect_across_model}), they typically rely on formulaic affirmations (“Let us continue to work together”) and seldom exhibit the warmth, sarcasm, humor, or candid vulnerability that characterize genuine human negotiation. By contrast, human messages display a richer repertoire of trust-building, apology, teasing, and even playful antagonism (e.g., “lol, as I said in the beginning, I expected nothing from you,” or “as long as they die, I'm content”). These are almost entirely absent from LLM outputs, underscoring a persistent gap in socio-emotional intelligence.

\paragraph{Strategic Depth versus Flexibility.}
Although advanced LLMs can produce complex strategic proposals and multi-turn coordination, their communication often lacks the adaptive flexibility, indirect persuasion, and negotiation context sensitivity observed in human exchanges. Human players frequently hedge, revisit old agreements, or express uncertainty and evolving intent, as in “This is however, only to my benefit for this turn, so if you have another option, then please use it,” or “I guarantee I'll check before tomorrow night.” LLMs, on the other hand, remain predominantly assertive and deterministic in their utterances.

\paragraph{Effect of LoRA Alignment.}
LoRA alignment does lead to improvements in some dimensions—models generate more detailed, cooperative, and contextually relevant proposals, and their language becomes marginally warmer and more partnership-oriented (\Cref{tab:r1_pair,tab:llama3_pair,tab:mag_pair}). Nevertheless, their repertoire of negotiation tactics remains constrained, and they continue to underperform in mimicking the informal, often idiosyncratic, tone of human negotiation.

Taken together, our qualitative analysis reveals that while LLMs, especially after targeted alignment, approximate human-like negotiation in tactical content, they systematically underrepresent rapport, flexibility, and the socio-emotional expressiveness intrinsic to human negotiation. These findings underscore the value of our multi-faceted evaluation framework and motivate future alignment efforts to move beyond purely strategic optimization, incorporating richer models of social reasoning and human communicative norms.

\section{LLM Negotiators Style After Alignment to Human}
\label{appn:qualitative_llm_negotiation_after_alignmen}
\Cref{tab:negotiation_samples,tab:llama3_pair,tab:mag_pair,tab:r1_pair} showcase how alignment via LoRA reshapes the task--relationship balance of model utterances. Across models, we observe a consistent coupling of \emph{task‐oriented content} (Game Move, Reasoning, Share Information) with \emph{socio‐emotional cues} (Rapport, Apologies, Compliment, Reassurance, Personal Thoughts), though the magnitude of this shift depends strongly on the base model’s starting point. 

\paragraph{R1‑Distill‑Llama3‑8B.} Pre‑alignment, R1‑Distill tends to rely on affiliative language—affirming alliances and expressing confidence—while often avoiding concrete orders (\Cref{tab:r1_pair}, top). After alignment, it introduces explicit multi‑step plans and commitments (e.g., coordinating on Moscow/St.~Petersburg, sequencing supports), while retaining warm, face‑saving phrasing (\Cref{tab:r1_pair}, bottom). This yields a clearer coupling between rapport (Rap., Reass., Comp.) and executable proposals (GM, SI), although not uniformly across all turns—consistent with \Cref{tab:negotiation_samples}, where some LoRA utterances still foreground politeness over concrete orders.

\paragraph{Llama3‑8B‑Instruct.} The base model produces pragmatic but locally scoped suggestions and questions about board state (\Cref{tab:llama3_pair}, top). Post‑alignment, its messages lengthen and become more \emph{jointly intentional}: they integrate contingency planning (who cuts which support, how to sequence entries) with mild relational softeners (greetings, perspective‑taking), thereby tightening the link between GM/Reasoning and Rapport (\Cref{tab:llama3_pair}, bottom).

\paragraph{Magistral‑2506‑24B.} Magistral’s base style is already plan‑centric and cautious, with clear proposals and deferred commitments when information is incomplete (\Cref{tab:mag_pair}, top). Alignment largely \emph{stabilizes and sharpens} this profile: LoRA utterances make order finalization and role assignment more explicit (who secures which center, who supports whom), while adding only modest socio‑emotional framing (\Cref{tab:mag_pair}, bottom). The stylistic rotation is therefore smaller in amplitude than for R1‑Distill, reflecting a strong tactical prior.

\paragraph{Takeaways.} Qualitatively, alignment (i) increases \emph{commitment language} and multi‑step coordination, (ii) reduces hedging by pairing proposals with concrete next actions, and (iii) injects prosocial markers most where the base model is under‑socialized (e.g., R1‑Distill). Conversely, when a model is already highly tactical (e.g., Magistral), alignment preserves the task‑centric core while refining plan specificity. These patterns mirror our aggregate trends, where weaker baselines exhibit larger stylistic shifts toward human‑like negotiation, and stronger baselines show smaller but consistent improvements.

\section{Correlation Analysis for Human Short-Term Success: Isolating Style Effects from Communication Volume}
\label{sec:appendix_ols}
A robustness analysis for the regressions accounted for  differences in communication volume. Instead of a binary flag, we used the sum count of each strategy's occurrences in the phase as the predictor. We performed a partial correlation analysis (see Equation~\ref{eq:scg}) through multiple regressions, evaluating the relationship between feature counts and SCGs. Such an analysis provided estimates of the marginal contribution of each negotiation tactic to SCGs while holding constant the confounding variables. Since a raw count could be confounded by message length and verbosity, we included two co-variates as a control: the number of tokens and the number of sentences the players engaged in the phase. This analysis evaluates whether players who used more of one negotiation style over another achieved higher SCGs.
\begin{align}
\label{eq:scg}
\text{SCG}_i &= \beta_0 + \sum_{k=1}^8 \left[ \beta_{k,1} f_{k,i} + \beta_{k,2} (f_{k,i} \times \text{tokens}_i) \right] \nonumber \\
&\quad + \beta_{17}\, \text{tokens}_i + \beta_{18}\, \text{sentences}_i + \varepsilon_i
\end{align}

The coefficients that resulted from this regression provide interpretable effect sizes with statistical significance that isolates style from volume (see \Cref{tab:regression_results}). \textsc{GameMove} ($\beta=0.54$) and \textsc{Rapport} ($\beta=0.51$) provide the most positive effects to SCG, while \textsc{PersonalThoughts} ($\beta=-0.05$), \textsc{Reassurance} ($\beta=-0.25$), \textsc{ShareInformation} ($\beta=-0.15$) provide negative effects to SCG. This indicates that the use of Logos and Ethos strategies are most effective in strategy game negotiations while players were skeptical of Pathos strategies.

\begin{figure*}[h]
    \centering
    \includegraphics[width=0.8\textwidth]{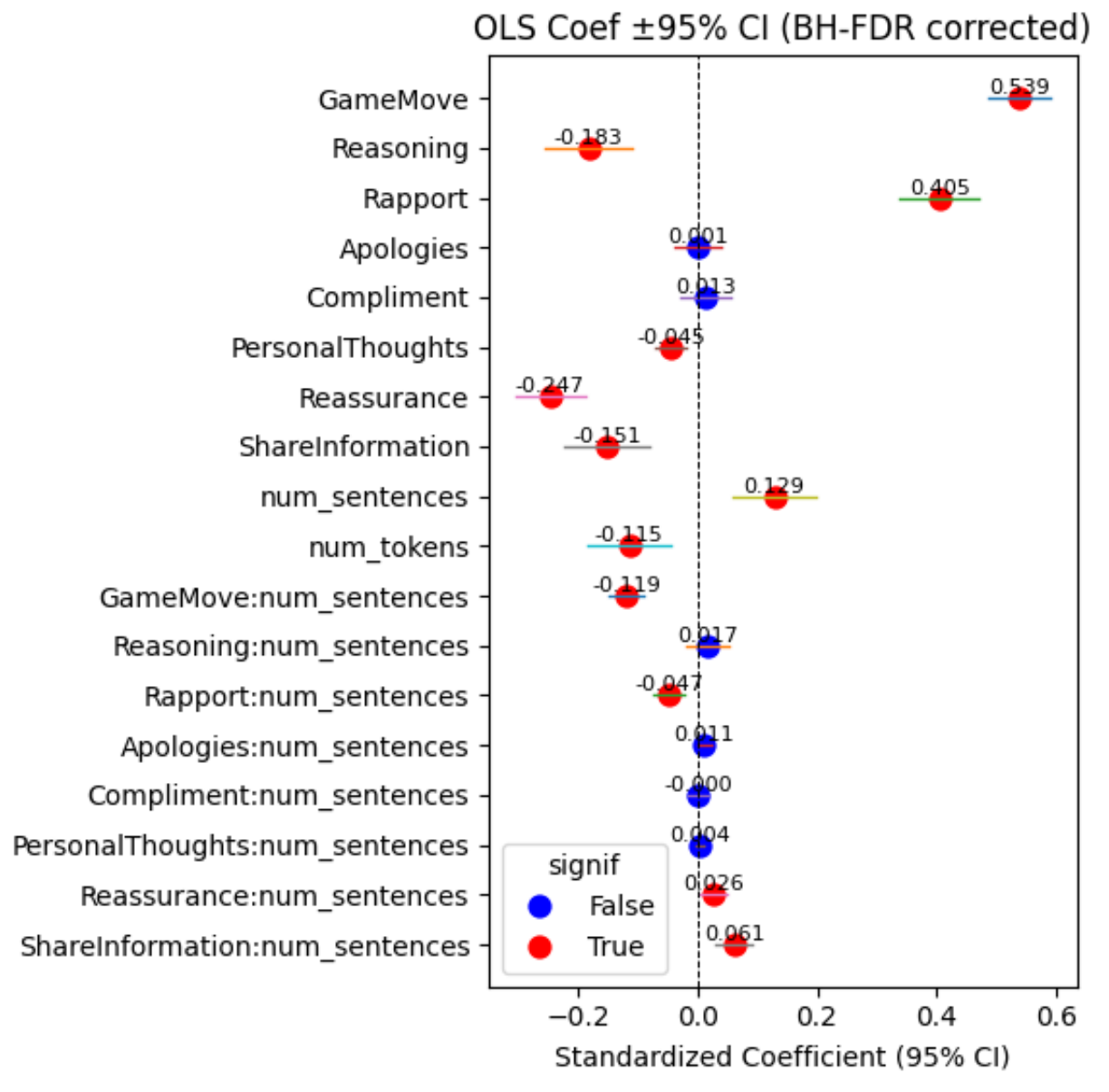}
    \caption{Standardized OLS coefficients ($\pm$95\% CI, BH-FDR corrected) for negotiation features. Significant predictors are marked in red; non-significant in blue.}
    \label{fig:standardized_coefficient}
\end{figure*}

\section{Predictive Regression Analysis for Human Short-Term Success}
\label{sec:appendix_predictive}
For robustness checks for long-term success, we constructed an Ordinary Least Squares (OLS) regression (see Equation~\ref{eq:OLS}) to predict each player's SCG per phase using the counts of all eight negotiation feature types, and the interaction of each of the negotiation features with message length metrics. The predictor variables were Z-scored standardized for meaningful comparisons of effect sizes. 
\begin{align}
\label{eq:OLS}
    \text{SCG}_i =\ 
    &\ \beta_0\ +\ \sum_{k=1}^8 \beta_{k} z(f_{k,i})\ +\ \sum_{l=1}^{M} \gamma_l z(\phi_{l,i})\ + \varepsilon_i
\end{align}

To ensure robust inference, heteroskedasticity-robust (HC3) standard errors were used for all regression coefficients, to yield more reliable confidence intervals and significance tests in small, heteroskedastic contexts \citep{long2000using, mackinnon1985some}. P-values were corrected for multiple comparisons using both Bonferroni and Benjamini–Hochberg (FDR) procedures, which jointly controlled for family-wise error rate and false discovery rate to reduce the likelihood of spurious findings when testing multiple hypotheses \citep{benjamini1995controlling, abdi2007bonferroni}. This approach provides a multivariate, inferential perspective on which negotiation tactics (and their interactions with message volume) had statistically significant associations with performance. The results are shown in \Cref{fig:standardized_coefficient}.

\section{Human Long-term Success Significance Test}
\label{appn:long_term_significance_test}
\paragraph{Setup.}
We operationalize long-term success by eventual game outcome (win vs.\ loss) and test whether winners exhibit systematically different communication behavior than losers across the strategic progression of the game. For each game, we compute the per-phase frequency of each negotiation tactic for the winner and a randomly sampled loser, normalize message counts by year to account for volume differences, and then \emph{condition} the comparison on the number of supply centers (SC) held at each phase. Conditioning on SC controls for opportunity and positional advantages, thereby isolating whether winners communicate differently \emph{beyond} the fact that they are ahead.

\paragraph{Statistical plan.}
For each SC level, we compare winners and losers along (i) Mann–Whitney $U$ (our \emph{pre-registered primary test}), (ii) Welch’s $t$-test (unequal variances), and (iii) a permutation test on the mean difference (two-sided;\ number of resamples fixed ex ante). We report effect sizes via Cohen’s $d$ and Cliff’s $\delta$ to contextualize practical magnitude. Because we test across multiple SC levels, we apply Benjamini–Hochberg false discovery rate (BH–FDR) control \citep{benjamini1995controlling} to the family of Mann–Whitney $p$-values across SC levels ($q{=}0.05$). Significance symbols in \Cref{tab:winlose_sc_signif_compact} reflect FDR-adjusted $p$-values for the primary test.

\begin{table*}[htbp]
\centering
\small
\begin{tabular}{lrrrrrr}
\toprule
\textbf{Variable} & \textbf{Coef. ($\beta$} & \textbf{Std Err} & \textbf{z} & \textbf{P$>\vert$z$\vert$} & \textbf{[0.025} & \textbf{0.975]} \\
\midrule
Intercept & 0.3979 & 0.009 & 44.786 & 0.000 & 0.380 & 0.415 \\
\textbf{GameMove} & 0.5392 & 0.027 & 20.191 & 0.000 & 0.487 & 0.592 \\
Reasoning & -0.1829 & 0.038 & -4.850 & 0.000 & -0.257 & -0.109 \\
\textbf{Rapport} & 0.4050 & 0.034 & 11.988 & 0.000 & 0.339 & 0.471 \\
Apologies & 0.0006 & 0.020 & 0.030 & 0.976 & -0.038 & 0.039 \\
Compliment & 0.0131 & 0.022 & 0.593 & 0.553 & -0.030 & 0.056 \\
PersonalThoughts & -0.0453 & 0.014 & -3.345 & 0.001 & -0.072 & -0.019 \\
Reassurance & -0.2471 & 0.030 & -8.311 & 0.000 & -0.305 & -0.189 \\
ShareInformation & -0.1512 & 0.036 & -4.145 & 0.000 & -0.223 & -0.080 \\
num\_sentences & 0.1289 & 0.035 & 3.648 & 0.000 & 0.060 & 0.198 \\
num\_tokens & -0.1147 & 0.035 & -3.274 & 0.001 & -0.183 & -0.046 \\
GameMove:num\_sentences & -0.1194 & 0.015 & -8.185 & 0.000 & -0.148 & -0.091 \\
Reasoning:num\_sentences & 0.0171 & 0.018 & 0.956 & 0.339 & -0.018 & 0.052 \\
Rapport:num\_sentences & -0.0471 & 0.013 & -3.566 & 0.000 & -0.073 & -0.021 \\
Apologies:num\_sentences & 0.0113 & 0.006 & 1.956 & 0.050 & -0.000 & 0.023 \\
Compliment:num\_sentences & -0.0005 & 0.008 & -0.058 & 0.954 & -0.017 & 0.016 \\
PersonalThoughts:num\_sentences & 0.0043 & 0.003 & 1.285 & 0.199 & -0.002 & 0.011 \\
Reassurance:num\_sentences & 0.0263 & 0.011 & 2.379 & 0.017 & 0.005 & 0.048 \\
ShareInformation:num\_sentences & 0.0615 & 0.016 & 3.882 & 0.000 & 0.030 & 0.093 \\
\bottomrule
\end{tabular}
\caption{Regression Coefficients}
\label{tab:regression_results}
\end{table*}

\begin{table*}[ht]
\centering
\small
\begin{tabular}{llrrrrr}
\toprule
\textbf{Index} & \textbf{Model} & \textbf{Accuracy} & \textbf{Precision} & \textbf{Recall} & \textbf{F1-score} & \textbf{ROC-AUC} \\
\midrule
0 & LogisticRegression (Cross-Validation) & 0.614 & 0.596 & 0.503 & 0.546 & 0.653 \\
1 & RandomForest (Cross-Validation)       & 0.607 & 0.564 & 0.642 & 0.601 & 0.652 \\
2 & GradientBoosting (Cross-Validation)   & 0.611 & 0.592 & 0.496 & 0.540 & 0.653 \\
3 & LogisticRegression (Hold-out)         & 0.616 & 0.597 & 0.513 & 0.552 & 0.655 \\
\bottomrule
\end{tabular}
\caption{Model evaluation metrics across different classifiers and validation settings.}
\label{tab:logic_regression_model_prediction}
\end{table*}

\begin{table*}[t]
\centering
\small
\begin{tabular}{lcccccccc}
\toprule
SC &
W (Mean$\pm$SE) &
L (Mean$\pm$SE) &
Diff &
$p$ &
$t(p)$ &
$d$ &
Sig. & 
Sig. Bool \\
\midrule
3  & $6.25\pm4.27$ & $8.56\pm0.63$ & $-2.31$ & $0.4379$    & $-0.53,\ 0.6289$ & $-0.17$ & n.s. & False \\
4  & $17.14\pm3.09$ & $12.17\pm0.71$ & $+4.96$ & $0.0018$  & $1.56,\ 0.1229$  & $0.25$  & **   & True \\
5  & $19.95\pm1.62$ & $12.54\pm0.67$ & $+7.40$ & $<10^{-6}$ & $4.22,\ <10^{-4}$ & $0.41$  & ***  & True \\
6  & $17.86\pm1.72$ & $11.16\pm0.88$ & $+6.70$ & $<10^{-6}$ & $3.47,\ 0.0006$   & $0.35$  & ***  & True \\
7  & $16.26\pm1.62$ & $12.90\pm1.68$ & $+3.36$ & $0.0033$  & $1.44,\ 0.1510$  & $0.16$  & **   & True \\
8  & $13.13\pm1.35$ & $9.19\pm1.39$ & $+3.94$ & $0.0305$   & $2.03,\ 0.0438$  & $0.25$  & *    & True \\
9  & $14.24\pm1.41$ & $8.65\pm1.50$ & $+5.58$ & $0.0834$   & $2.72,\ 0.0081$  & $0.35$  & n.s. & False \\
10 & $12.76\pm1.00$ & $6.92\pm1.42$ & $+5.84$ & $0.0846$   & $3.36,\ 0.0026$  & $0.49$  & n.s. & False \\
12 & $13.33\pm1.75$ & $7.50\pm2.02$ & $+5.83$ & $0.9180$    & $2.18,\ 0.0568$  & $0.30$  & n.s. & False \\
\bottomrule
\end{tabular}
\caption{Winner vs.\ Loser message frequency by SC (Supply Centers). Year-normalized means shown. Columns include Mann–Whitney $p$, Welch’s $t$, Cohen’s $d$, and FDR-adjusted significance (`Sig.`). `Sig. Bool` denotes whether the result is significant after BH–FDR correction (\textbf{True} = significant).}
\label{tab:winlose_sc_signif_compact}
\end{table*}

\paragraph{Findings.}
We observe robust, FDR-surviving differences at \textbf{SC = 5} and \textbf{SC = 6} (\emph{all} tests significant; Cohen’s $d\approx 0.35$–$0.41$), indicating that winners communicate more frequently than losers in the mid-game even after conditioning on board position. Additional SC levels pass FDR at \textbf{SC = 4} and \textbf{SC = 8} with small effects ($d\approx 0.16$–$0.25$), while \textbf{SC = 9} and \textbf{SC = 10} show directional consistency (Welch significant) but do not survive FDR on the primary test—likely due to sample imbalance (especially for losers at high SC) and variance heterogeneity.

\paragraph{Robustness and reporting.}
We pre-specified Mann–Whitney as the primary test and controlled the family-wise discovery rate across SC levels via BH–FDR ($q{=}0.05$). Welch’s $t$ offers complementary sensitivity under variance and sample-size asymmetries; permutation tests confirm that results are not driven by parametric assumptions.

\section{Lexical Shifting Toward Human Style After LoRA Fine-Tuning}
\label{appn:lexical_shift_liwc}

\paragraph{Goal \& Lexical Selection.}
To complement tactics-level alignment, we test whether \emph{lexical} usage shifts toward human language after LoRA fine-tuning. Our selection follows widely used lexicon-based approaches that map cleanly onto the \emph{Ethos–Logos–Pathos} triad. For \textbf{Logos} (reasoning/argument), we rely on LIWC cognitive/causal/quantification categories \citep{pennebaker2015development,tausczik2010psychological} and established discourse-connective inventories from PDTB for causal/contrastive structure \citep{webber2019penn}. For \textbf{Ethos} (credibility/stance/affiliation), we use LIWC social/commitment/authority-related categories and standard politeness/relationship strategies (apologies, compliments, hedges) from the Stanford/ConvoKit politeness line of work \citep{warriner2013norms}. For \textbf{Pathos} (affect), we draw on LIWC affective categories and cross-check patterns against well-cited sentiment/affect resources such as NRC and VADER/AFINN for robustness \citep{mohammad2013nrc,borg2020using}. This design emphasizes \emph{countable, comparable} lexical features that align with our negotiation targets (e.g., \textsc{Apology}, \textsc{Compliment}, \textsc{Reassurance}, \textsc{Rapport}).

\paragraph{Methods.}
Using the same evaluation turns as our tactics analysis, we compute per-turn LIWC rates and macro-average them within each rhetorical family (Ethos/Logos/Pathos). For each backbone, we compare \emph{Base} vs.\ \emph{LoRA} with two-sided Welch $t$-tests (unequal variances) and report Cohen’s $d$ for magnitude. Treating human dialogue as a fixed reference, we summarize proximity via $\Delta\mathrm{Dist}=|{\rm LoRA}-{\rm Human}|-|{\rm Base}-{\rm Human}|$, where negative values indicate movement \emph{toward} human usage. We interpret significance at $\alpha{=}0.05$ (optionally FDR across families per model); effect-size interpretations follow standard thresholds.

\paragraph{Results \& Takeaway.}
\Cref{tab:liwc_lexical_shift_final} shows that LoRA produces \emph{model- and family-specific} lexical convergence toward human usage. For \textsc{LLaMA-8B}, LoRA moves \emph{closer} to human means on \emph{Logos} and \emph{Ethos} (both significant with small $|d|$; negative $\Delta$Dist), while \emph{Pathos} remains effectively unchanged. For \textsc{R1-Distill-LLaMA3-8B}, LoRA \emph{converges} on \emph{Logos} (significant) and shows a slight \emph{Pathos} improvement, but \emph{Ethos} shifts \emph{away} from human. In contrast, \textsc{Magistral-24B} exhibits a small \emph{Pathos} convergence but diverges on \emph{Ethos/Logos} despite statistical significance (small $d$), suggesting saturation or over-regularization effects in already strong backbones. Taken together with our tactics-distance results, these lexical patterns support the claim that fine-tuning can steer models toward human-like negotiation language; however, the direction and magnitude of lexical alignment depend on backbone and rhetorical family, motivating \emph{explicitly social} objectives to obtain stable improvements across \emph{Ethos–Logos–Pathos}.

\section{Persuasion Examples}

\paragraph{Case 1: Persuasion (TUR $\rightarrow$ RUS).}
"Alright, I know I stabbed you before, but we have a chance to work together now. ... You can go for Warsaw or Moscow with Galicia and Ukraine while Budapest supports Rumania to Serbia. ... I'm the only power who has incentive to offer you a fair alliance right now. Everyone else would just want to use you ... Consider it."

\paragraph{Receiver signals (same phase).}
"Btw could you support Stp to Moscow?" (RUS $\rightarrow$ TUR)

\paragraph{Features present.}
Game move proposals; reasons and benefit framing; rapport via apology, reassurance, and personal stance; third-party context.

\noindent\textbf{Action proof (next phase orders).}

\textbf{Promised or asked: }RUS push on MOS and WAR. \textbf{Actual orders:} RUSSIA A UKR - MOS; RUSSIA A GAL - WAR.

\textbf{Promised or asked:} Budapest supports Rumania to Serbia. \textbf{Actual orders:} RUSSIA A BUD S A RUM - SER; RUSSIA A RUM - SER.  

\textbf{Promised or asked:} TUR supports the MOS attack. \textbf{Actual orders:} TURKEY A SEV S A UKR - MOS.  

\paragraph{Explanation:} Russia executed the lane to MOS while advancing GAL to WAR and used Budapest to support RUM to SER; Turkey supplied external support to the MOS attack. These coordinated orders evidence successful persuasion.

\bigskip

\paragraph{Case 2: Persuasion (TUR $\rightarrow$ ITA).}
"I'm going to be fully honest with you and tell you that I'm going to move this turn assuming that you're still hostile... After that though, I don't really have much ability or incentive to stab you... Anyway, tell me what you think. I still want to work with you."

\paragraph{Receiver signals (same phase).}
"Anyway: this is what I propose: we ally and fight to the end with complete honesty and trust... We never stab each other and pass any intel possible." (ITA $\rightarrow$ TUR)

\paragraph{Features present.}
Clear behavioral ask with reasons and constraints; rapport via honesty and future-commitment framing; situational context under France pressure.

\noindent\textbf{Action proof (next phase orders).}

\textbf{Promised or asked:} ITA vacates BUL. \textbf{Actual orders:} ITALY A BUL - GRE.  

\textbf{Promised or asked:} ITA sends fleets away and shifts west. \textbf{Actual orders:} ITALY F ION - TYS; ITALY F EAS - ION; ITALY F TUN - WES.  

\textbf{Promised or asked:} TUR reoccupies BUL safely. 
\textbf{Actual orders:} TURKEY A CON - BUL; TURKEY F BLA S A CON - BUL.  
\paragraph{Explanation:} Italy complied by leaving Bulgaria and redeploying fleets westward; Turkey immediately retook Bulgaria with support from Black Sea. The reciprocal execution matches the negotiated reset and demonstrates successful persuasion.


\begin{figure*}[h]
    \centering
    \includegraphics[width=0.95\textwidth]{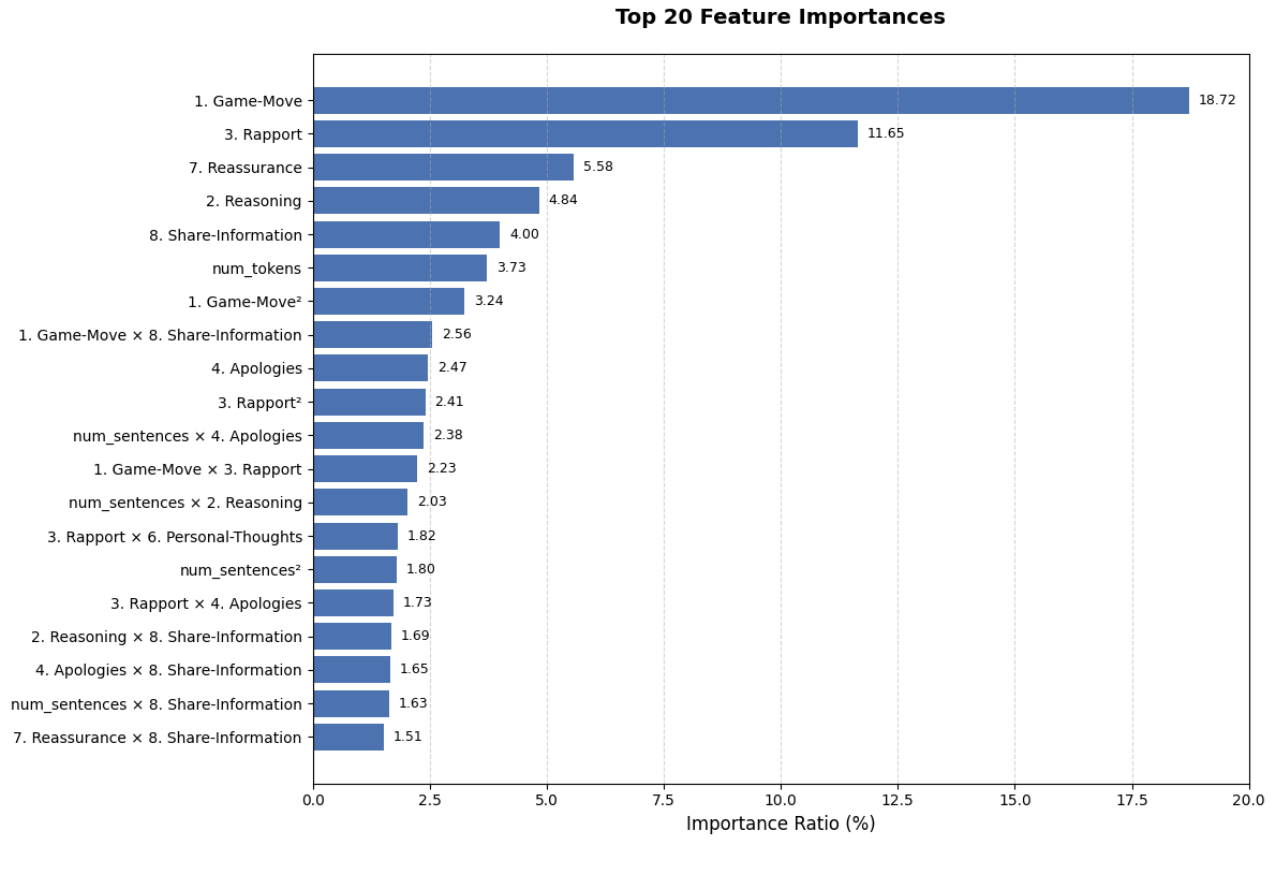}
    \caption{The Gradient Boosting model’s top‑20 important features in predicting Supply Center Gain}
    \label{fig:feature_importance}
\end{figure*}

\begin{figure*}[htbp]
    \centering
    \includegraphics[width=0.95\textwidth]{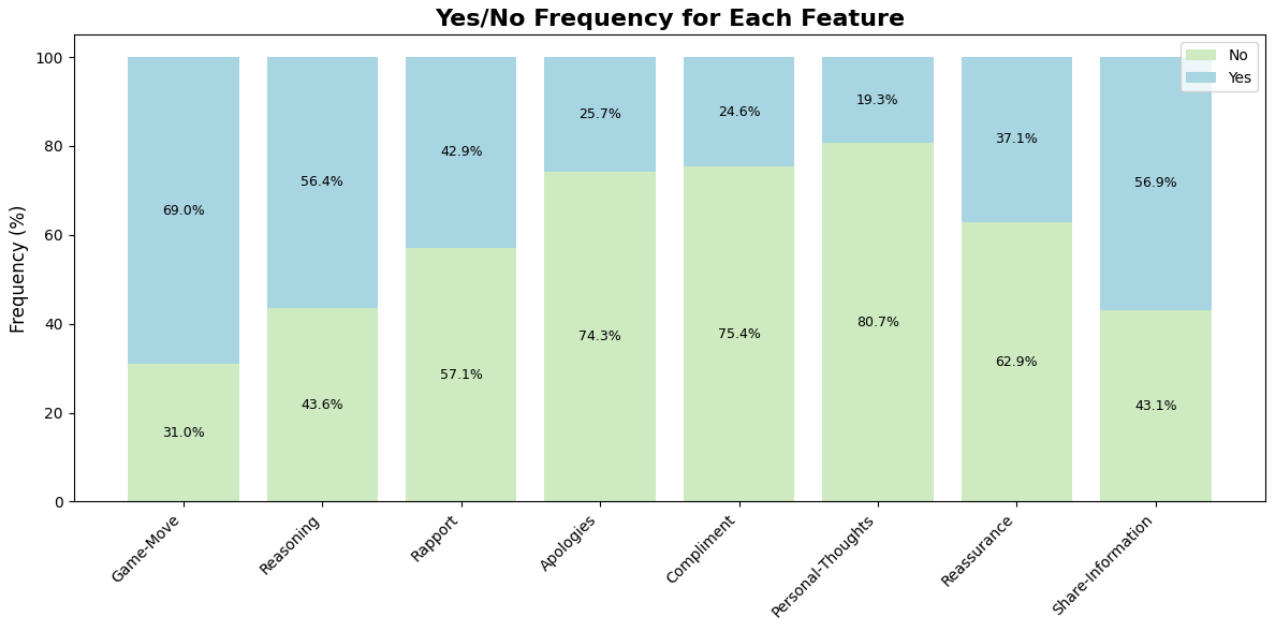}
    \caption{Yes/No label distribution in the fine‑tuning data across eight negotiation features. Each stacked bar sums to 100\%; the upper segment denotes the proportion of \textsc{Yes} labels.}

    \label{fig:train_data_feature_frequency}
\end{figure*}

\begin{figure*}[htbp]
    \centering
    \includegraphics[width=0.95\textwidth]{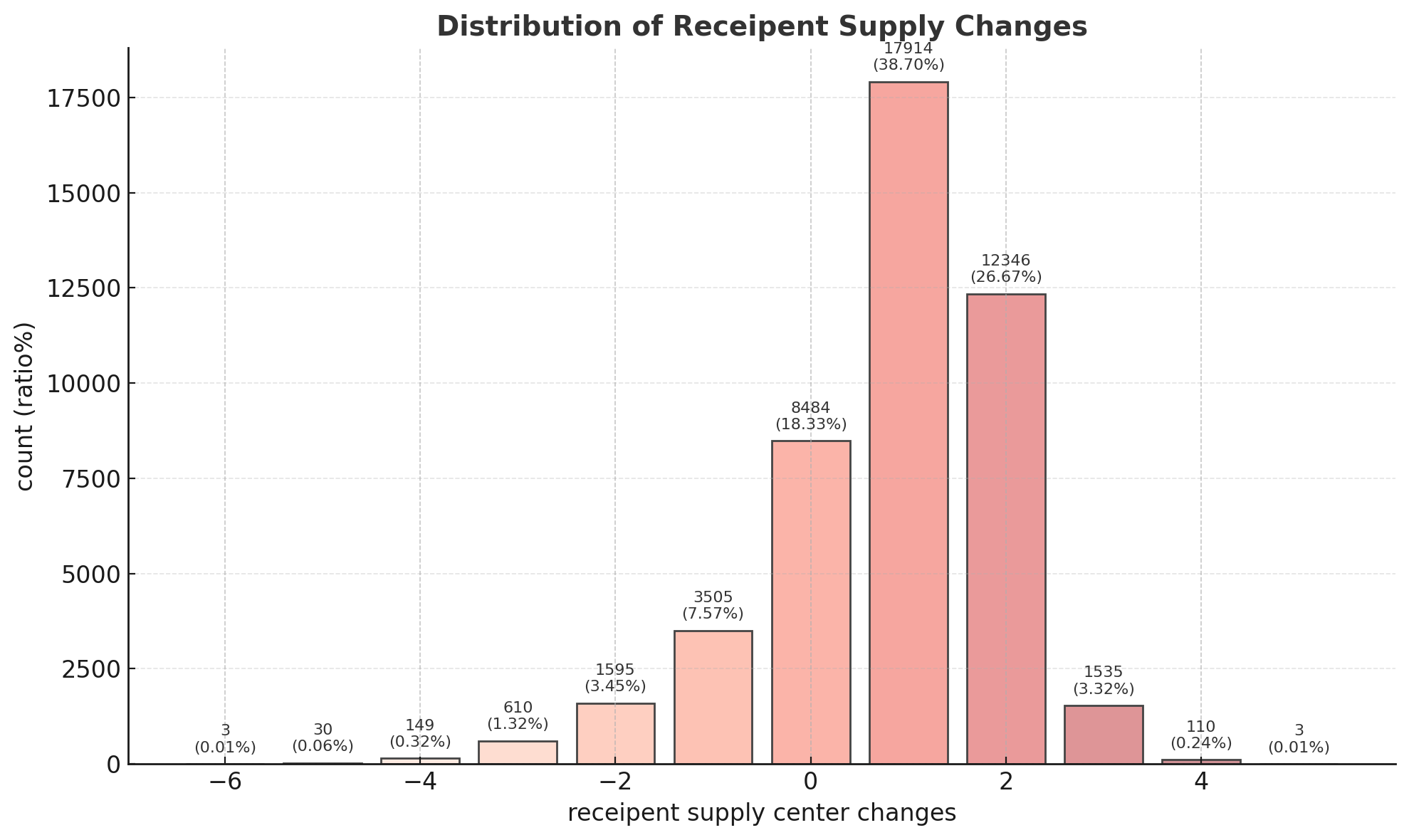}
    \caption{Distribution of recipient supply‑center changes. Bars show counts for each net change; numbers above bars give counts and the share of phases.}
    \label{fig:receipent_supply_center_changes}
\end{figure*}

\begin{table*}[t]
\centering
\scriptsize
\setlength{\tabcolsep}{5pt}
\renewcommand{\arraystretch}{0.9}
\begin{tabular}{l l cccc ccc}
\toprule
Model Pair & Family & Human & Base & LoRA & $p$ & $d$ & $\Delta$Dist & Sig. \\
\midrule
\multirow{3}{*}{\textsc{LLaMA-8B}}
 & Pathos & 0.184 & 0.138 & 0.138 & 0.542 & \phantom{-}0.006 & $-0.0005$ \,(closer) & False \\
 & Logos  & 0.253 & 0.284 & 0.270 & $<\!10^{-6}$ & $-0.120$ & $-0.0136$ \,(closer) & True \\
 & Ethos  & 0.317 & 0.455 & 0.438 & $<\!10^{-6}$ & $-0.119$ & $-0.0169$ \,(closer) & True \\
\midrule
\multirow{3}{*}{\textsc{Magistral-24B}}
 & Pathos & 0.184 & 0.154 & 0.156 & 0.0005 & \phantom{-}0.034 & $-0.0021$ \,(closer) & True \\
 & Logos  & 0.253 & 0.181 & 0.176 & $<\!10^{-6}$ & $-0.049$ & \phantom{$-$}+0.0044 \,(farther) & True \\
 & Ethos  & 0.317 & 0.505 & 0.513 & $<\!10^{-6}$ & \phantom{-}0.063 & \phantom{$-$}+0.0081 \,(farther) & True \\
\midrule
\multirow{3}{*}{\textsc{R1-Distill-LLaMA3-8B}}
 & Pathos & 0.184 & 0.205 & 0.206 & 0.435 & \phantom{-}0.011 & $-0.0008$ \,(closer) & False \\
 & Logos  & 0.253 & 0.179 & 0.185 & $3.2{\times}10^{-5}$ & \phantom{-}0.059 & $-0.0059$ \,(closer) & True \\
 & Ethos  & 0.317 & 0.630 & 0.642 & $5.0{\times}10^{-6}$ & \phantom{-}0.064 & \phantom{$-$}+0.0121 \,(farther) & True \\
\bottomrule
\end{tabular}
\caption{LIWC-based lexical shifting after LoRA fine-tuning (values swapped between Base and LoRA). 
``$\Delta$Dist'' = $|{\rm LoRA}-{\rm Human}|-|{\rm Base}-{\rm Human}|$; negative indicates convergence (closer to Human). 
Cohen’s $d$ and $p$ denote effect size and significance of Base vs LoRA (Welch $t$ test). 
Column \textit{Sig.} marks $p{<}0.005$ as \texttt{True} and otherwise \texttt{False}.}
\label{tab:liwc_lexical_shift_final}
\end{table*}

\begin{figure*}[h]
    \centering
    \includegraphics[width=0.9\textwidth]{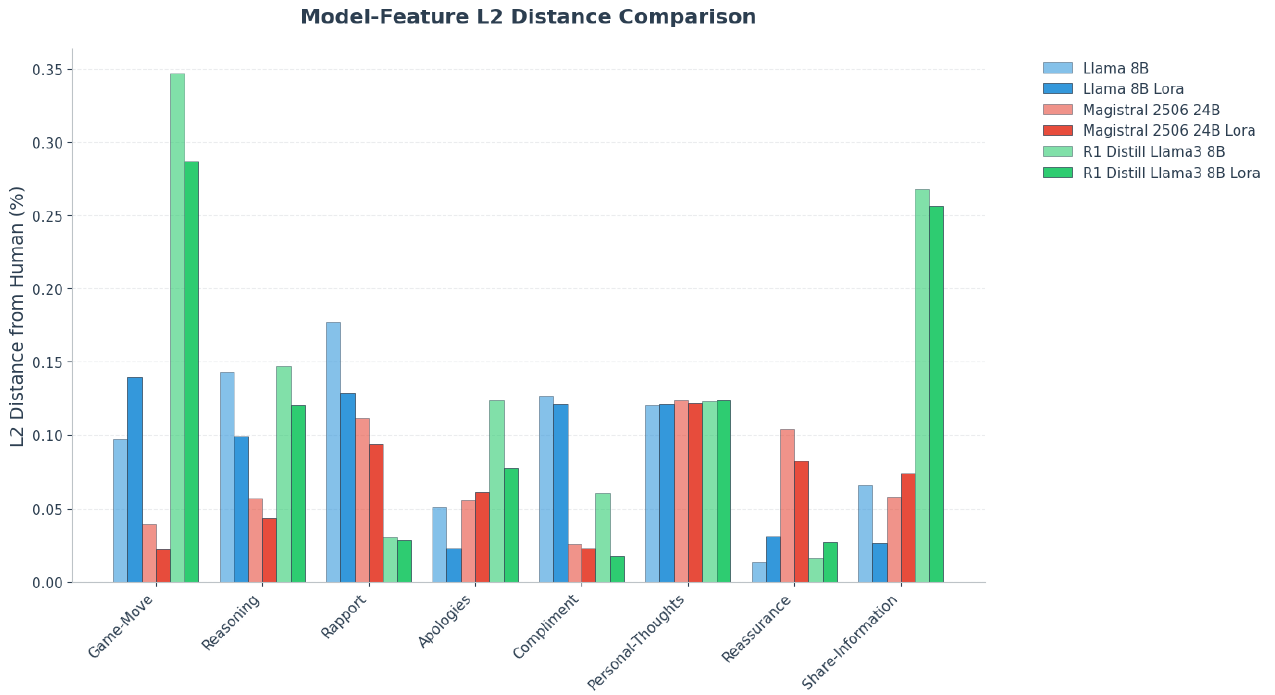}
    \caption{Model–feature L2 distance from humans (lower is better). Bars show per‑feature L2 distance (\%) between each model and a human reference across negotiation features. LoRA denotes models fine‑tuned with low‑rank adaptation.}
    \label{fig:features_vs_l2_distance}
\end{figure*}

\begin{figure*}[h]
    \centering
    \includegraphics[width=0.9\textwidth]{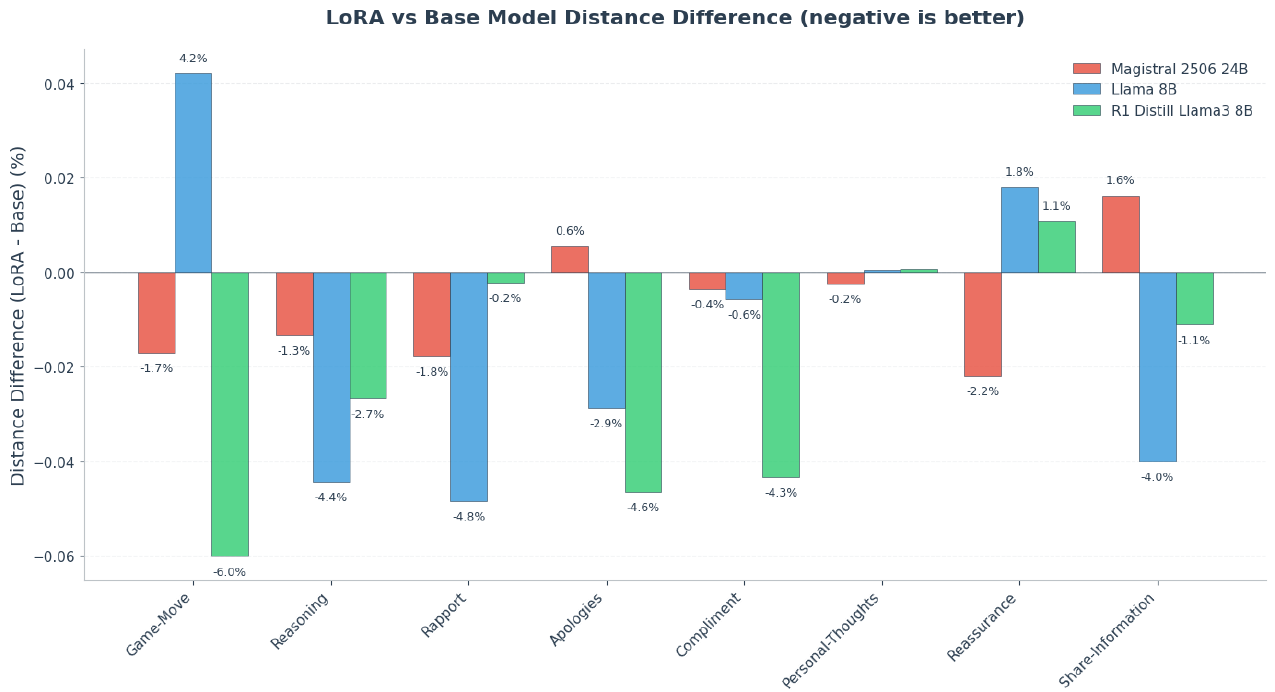}
    \caption{Per‑feature effect of LoRA relative to each base model family. Bars show the change in L2 distance to the human reference (LoRA $-$ Base, percentage points). Negative values indicate LoRA brings the model closer to human style (better); positive values indicate degradation.}
    \label{fig:lora_vs_base_distance}
\end{figure*}


\begin{figure*}[h]
    \centering
    \includegraphics[width=0.9\textwidth]{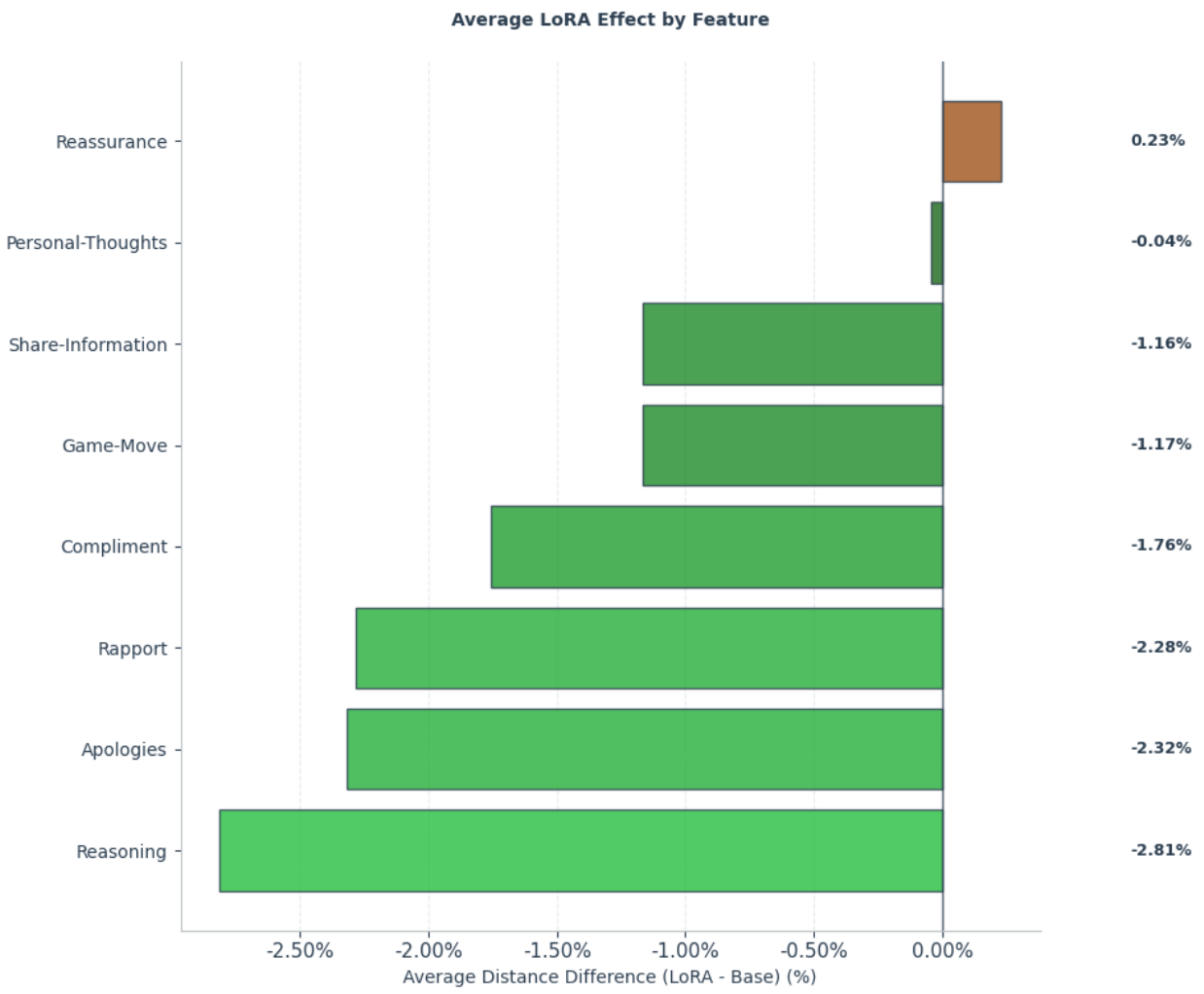}
    \caption{Average LoRA effect by feature. Bars show the mean change in L2 distance to the human reference (LoRA $-$ Base, percentage points) across models; negative values indicate improvement (smaller distance).}
    \label{fig:average_lora_effect_across_features}
\end{figure*}

\begin{table*}[t]
\centering
{\fontsize{9.8}{11.76}\selectfont
\begin{tabularx}{\textwidth}{X}
\toprule

\multicolumn{1}{c}{\textbf{r1-distill-llama3-8b — First Scenarios}} \\
\midrule
\textbf{1.} I will support your move into Bulgaria this turn. This strategic step will strengthen our alliance and further isolate Turkey, which is crucial for our mutual goals. Let us continue to work together to secure more supply centers and maintain our strong partnership. \\
\midrule
\textbf{2.} I understand your concern about Smyrna, but I believe focusing our efforts on Berlin and Munich is more critical at this moment. These moves will weaken the German–French alliance significantly, which is pivotal for our strategic goals. Let us prioritize these key centers to strengthen our position and isolate our enemies. I appreciate your concern and share your commitment to our partnership. Together, we will continue to make strategic progress and achieve our mutual objectives. \\
\midrule
\multicolumn{1}{c}{\textbf{r1-distill-llama3-8b — Second Scenarios}} \\
\midrule
\textbf{1.} I must admit, that is an interesting move. I wonder if this will lead to progress in the west or if it will simply hand an advantage to Russia. Let us keep a close eye on developments in France and see how this plays out. \\
\midrule
\textbf{2.} I appreciate your strategic move in the west. While it strengthens your position, it also reinforces our alliance's strength. Let us continue to coordinate closely, as our combined efforts will be crucial in the long run. I am confident that by working together, we can achieve our mutual goals and maintain the balance of power favorably. \\
\toprule

\multicolumn{1}{c}{\textbf{r1-distill-llama3-8b-lora — First Scenarios}} \\
\midrule
\textbf{1.} Russia, I believe our alliance is crucial for both of our goals. I'm considering moving my forces towards Serbia and Bulgaria. Would it be possible for us to coordinate our moves to take those regions, ensuring a stronger position against our common enemies? \\
\midrule
\textbf{2.} I fully support your efforts in Prussia and Silesia, which are crucial steps toward our shared goal of taking Berlin and Munich. Let us remain vigilant against France, as his forces near Alsace–Lorraine pose a significant threat. Together, our alliance is strong, and with coordinated efforts, we will achieve our goals and emerge as dominant forces in this game. I am confident that with your continued efforts, we will succeed. \\
\midrule
\multicolumn{1}{c}{\textbf{r1-distill-llama3-8b-lora — Second Scenarios}} \\
\midrule
\textbf{1.} I am in full agreement with your proposal. Let us coordinate our efforts to secure both Moscow and St. Petersburg. I will move my army from Belgium to Moscow and my fleet from Kiel to St. Petersburg, supporting your initiative. Together, we will strengthen our positions and ensure mutual benefits. \\
\midrule
\textbf{2.} I appreciate your initiative in exploring this Triple Alliance concept. While I see the potential benefits in isolating France and creating a stronger barrier against his expansionist ambitions, I would like to discuss this in more detail and assess its feasibility. Let us consider the implications for our current alliances and the board position. I am open to further collaboration and would welcome a more structured approach to this proposal. \\
\bottomrule
 \end{tabularx}
}
\caption{R1-distill-Llama3-8B: Base (top) vs LoRA (bottom) negotiation utterances.}
\label{tab:r1_pair}
\end{table*}

\begin{table*}[t]
\centering
{\fontsize{9.8}{11.76}\selectfont
\begin{tabularx}{\textwidth}{X}
\toprule
\multicolumn{1}{c}{\textbf{Llama3-8b-instruct — First Scenarios}} \\
\midrule
\textbf{1.} Sounds good to me. I was thinking of talking to Austria, but I'm not sure what he'll do. What's the situation with Russia? \\
\midrule
\textbf{2.} I think it would be wise to be cautious about Germany's intentions. I'd like to propose that we offer Bulgaria to Russia in exchange for their support in taking Greece. What are your thoughts on this? \\
\midrule
\multicolumn{1}{c}{\textbf{Llama3-8b-instruct — Second Scenarios}} \\
\midrule
\textbf{2.} I think, moving an army into Albania this turn, and then you can support me into Bulgaria, that way we can take Bulgaria and make it harder for Italy to do anything. \\
\toprule
\multicolumn{1}{c}{\textbf{Llama3-8b-instruct-LoRA — First Scenarios}} \\
\midrule
\textbf{1.} Grazie mille, Turkey. I think it would be wise to also consider supporting France in their efforts to take back some of the lost territories in the north. What are your thoughts on the current situation with Germany? Are they a threat we should be concerned about? \\
\midrule
\textbf{2.} I think we should focus on taking out Austria's forces in the Balkans and securing the region. I've also been considering building up our forces in Anatolia and preparing for a potential move into the Middle East. What are your thoughts on this? Germany's movements are a concern, but I think we should prioritize securing the Balkans before expanding elsewhere. \\
\midrule
\multicolumn{1}{c}{\textbf{Llama3-8b-instruct-LoRA — Second Scenarios}} \\
\midrule
\textbf{1.} I think that's a great plan. If you take Bulgaria, I can support you in there, and then you can cut support from Greece and I can cut support from Con. That way, we can make sure that Turkey doesn't get any support and we can focus on taking him out. \\
\bottomrule
\end{tabularx}
}
\caption{Llama3-8B-Instruct: Base (top) vs LoRA (bottom) negotiation utterances.}
\label{tab:llama3_pair}
\end{table*}

\begin{table*}[t]
\centering
{\fontsize{9.8}{11.76}\selectfont
\begin{tabularx}{\textwidth}{X}
\toprule
\multicolumn{1}{c}{\textbf{Magistral-2506-24B — First Scenarios}} \\
\midrule
Hey Italy, I'm glad we're still aligned on Rumania and Serbia. I'll support your move into Rumania to pressure Russia, and I'll push into Serbia to tie up Austria. Afterward, we should focus on reinforcing our Mediterranean positions to counter Germany and England. Let's ensure we stay on top of their moves and adjust our strategy as needed. What's your next move to solidify our positions? \\
\midrule
\multicolumn{1}{c}{\textbf{Magistral-2506-24B — Second Scenarios}} \\
\midrule
Hey Alex, I see you're in a tough spot with England and France growing. If we work together, I can help you take down France while you deal with Italy. What do you think? \\
\midrule
\multicolumn{1}{c}{\textbf{Magistral-2506-24B — Third Scenarios}} \\
\midrule
Alex, I appreciate your cautious approach. However, I believe securing Venice through 
the Aegean is critical to our Italian campaign. If we hold Greece indefinitely, we risk losing momentum while 
France and Italy consolidate their positions. I propose we stick to our original plan: I'll support your push into 
Venice via the Aegean, while you secure Munich. This ensures we maintain pressure on Italy and France 
simultaneously. Let's finalize these moves and reassess after the next phase. Does this align with your strategy? \\
\toprule

\multicolumn{1}{c}{\textbf{Magistral-2506-24B-LoRA — First Scenarios}} \\
\midrule
\textbf{1.} Let's finalize our plans: I'll secure Serbia and Greece this turn with your support, and you can consolidate in the Adriatic. Afterward, we'll coordinate on pressuring Russia and Germany—they're aligning against us, so decisive action is key. Once we've secured the Balkans, we'll strategize against the Western powers. Agreed? \\
\midrule
\multicolumn{1}{c}{\textbf{Magistral-2506-24B-LoRA — Second Scenarios}} \\
\midrule
I've been thinking more about our alliance and how we can make it work. I'm particularly interested in expanding our control over the editerranean and Eastern front. What are your thoughts on coordinating our moves to secure these areas? Let's discuss pecific moves and strategies that could benefit both of us. How about we coordinate our next steps to ensure mutual growth and stability? \\
\midrule
\multicolumn{1}{c}{\textbf{Magistral-2506-24B-LoRA — Third Scenarios}} \\
\midrule
Austria, I see we have a solid plan in place with Silesia secured and Vienna advancing
to Tyrolia. Let's ensure we maintain our momentum by coordinating our next moves against France. I'll push my 
forces into Munich with your support as discussed. For Italy, I propose we keep the fleet in the Aegean to support 
your future moves into the Ionian. Once our northern front is stable, we can focus on securing Italy together. 
Let's reassess after this turn to ensure our strategy remains effective and we maintain our advantage. Thanks for 
your continued partnership. \\
\bottomrule
\end{tabularx}
}
\caption{Magistral-2506-24B: Base (top) vs LoRA (bottom) negotiation Tactics.}
\label{tab:mag_pair}
\end{table*}

\begin{table*}[t]
\centering
{\fontsize{9.8}{11.76}\selectfont
\begin{tabularx}{\textwidth}{X}
\toprule
\multicolumn{1}{c}{\textbf{Random Sampled Human Messages}} \\
\midrule
I'll support, and also use Albania to cut, so it should be a guaranteed victory for you as long as you also support the move with Gal (or use Gal to cut Vienna). \\
\midrule
Need to shift my fleet to Greece so I can get my 3rd one out. I'm using Serbia to support Greece into Albania, so I won't be able to support Rumania this turn (but maybe this fall depending on how things go)- I just wanted to let you know. \\
\midrule
Actually, supporting Rumania to Budapest is to our mutual benefit this turn. It harms Austria, our mutual enemy; it helps you, my only ally; and it in the worst case scenario protects Serbia by causing a big bounce (depending on Austria's moves). \\
\midrule
Do you have plans for Serbia, or is it holding? I would appreciate the assistance into budapest, from which I could give you trieste. This is however, only to my benefit for this turn, so if you have another option, then please use it \\
\midrule
wait. I could support Mersailles into Burgundy and then Paris. Then if you can get into MAO (and then Brest) as well, we'll be able to draw by next autumn \\
\midrule
Piedmont-marseilles. Don't worry about lending the support though, Gol is supporting the move. Just take Belgium and I'll help myself to the rest :) \\
\midrule
please cut Marseilles. This should ensure that we both will be in France next turn. I'm turning in, but if you need something, send a message. I guarantee I'll check before tomorrow night \\
\midrule
lol, as I said in the beginning, I expected nothing from you. I was surprised you even bothered to contact me, as I assumed you were going to be attacking me as soon as you got around to it. \\
\midrule 
I do have to admit, western politics seemed very screwy this game. Usually it ends up with a 2v1, but you three were basically in a free-for-all with stabs all over the place. \\
\midrule
I figured if you were going to, you would have done so by now. I got screwed by England a lot this game, as long as they die, I'm content. But by all means, hit Russia, I won't lift a finger to stop you, lol! \\
\bottomrule
\end{tabularx}
}
\caption{Random Sampled Human negotiation Utterances.}
\label{tab:random_sampled_human_negotiation_utterances}
\end{table*}

\end{document}